\documentclass[aps,pra,preprint,floatfix]{revtex4-1}

\usepackage{graphicx}
\usepackage{natbib}

\usepackage{amssymb, amsmath, amsfonts, wasysym, bm} 
\usepackage{color}          
\usepackage{float}          
\usepackage{mathrsfs}       
\usepackage{natbib}         
\usepackage{siunitx}        
\usepackage{subfigure}         
\usepackage{todonotes}      
\usepackage{units}          
\usepackage{url}
\usepackage{bigints}

\usepackage[colorlinks=true, 
            linkcolor=blue,
            urlcolor=blue,
            citecolor=blue,
            final,
            hypertexnames=false]{hyperref}
\usepackage{cleveref}

\newcommand{\commentout}[1]{}

\newcommand{\pp}   [2] {\frac{\partial #1}{\partial #2}}

\newcommand{\vvvert}   {|\kern-1pt|\kern-1pt|}

\newcommand{\diff} [1] {\operatorname{d}\!#1}
\newcommand{\dd}   [2] {\frac{\diff #1}{\diff #2}}

\newcommand{\Ssd}      {\mathcal{S}}                    
\newcommand{\mean} [1] {\ensuremath{\overline{#1}}}     
\newcommand{\fav}  [1] {\ensuremath{\widetilde{#1}}}    
\newcommand{\fluc} [1] {\ensuremath{#1'}}               
\newcommand{\ffluc}[1] {\ensuremath{#1''}}              
\newcommand{\grt}{\ensuremath{\gamma}}

\newcommand{\func} [2] {\ensuremath{#1 \! \left(#2\right)}}        
\newcommand{\tz}       {\ensuremath{t_0}}          

\newcommand{\lowM}     {L}                         
\newcommand{\cev}      {C}                         
\newcommand{\wall}     {\ensuremath{\mathrm{w}}}   
\newcommand{\awall}    {\ensuremath{\mathrm{aw}}}  
\newcommand{\inc}      {\ensuremath{\mathrm{inc}}} 

\newcommand{\Prt}{\ensuremath{\mbox{Pr}_t}}
\newcommand{\ueff}{\ensuremath{\tilde{u}_{\mathrm{eff}}}}
\newcommand{\vmod}{\ensuremath{\tilde{v}_{\mathrm{mod}}}}

\begin{document}

\title[Temporal Slow Growth DNS]{A Temporal Slow Growth Formulation
  for Direct Numerical Simulation of Compressible Wall--Bounded Flows}

\author{ Victor Topalian
  \footnote{Current affiliation: Esgee Technologies, Austin, Texas 78746, USA}}
\email{victopa@gmail.com}
\affiliation{Institute for Computational Engineering and Sciences,\\ 
The University of Texas at Austin, Austin, Texas 78712, USA}

\author{ Todd A.~Oliver }
\email{oliver@ices.utexas.edu}
\affiliation{Institute for Computational Engineering and Sciences,\\ 
The University of Texas at Austin, Austin, Texas 78712, USA}

\author{ Rhys Ulerich
  \footnote{Current affiliation: Two Sigma Investments, L.P., New York, New York 10013, USA}
}
\email{rhys@twosigma.com}
\affiliation{Institute for Computational Engineering and Sciences,\\ 
The University of Texas at Austin, Austin, Texas 78712, USA}

\author{Robert D.~Moser}
\email{rmoser@ices.utexas.edu}
\affiliation{Institute for Computational Engineering and Sciences,\\ 
The University of Texas at Austin, Austin, Texas 78712, USA}
\affiliation{Department of Mechanical Engineering,\\
The University of Texas at Austin, Austin, Texas 78712, USA}

\begin{abstract}
A new slow growth formulation for DNS of wall-bounded turbulent flow
is developed and demonstrated to enable extension of slow growth
modeling concepts to complex boundary layer flows.  As in previous
slow growth approaches, the formulation assumes scale separation
between the fast scales of turbulence and the slow evolution of
statistics such as the mean flow.  This separation enables the
development of approaches where the fast scales of turbulence are
directly simulated while the forcing provided by the slow evolution is
modeled.  The resulting model admits periodic boundary conditions in
the streamwise direction, which avoids the need for extremely long
domains and complex inflow conditions that typically accompany
spatially developing simulations.  Further, it enables the use of
efficient Fourier numerics.  Unlike previous
approaches~\citep{spalart1988direct,guarini2000direct}, the present
approach is based on a temporally evolving boundary layer and is
specifically tailored to give results for calibration and validation
of RANS turbulence models.  The use of a temporal homogenization
simplifies the modeling, enabling straightforward extension to flows
with complicating features, including cold and blowing walls.  To
generate data useful for calibration and validation of RANS models,
special care is taken to ensure that the mean slow growth forcing
is closed in terms of the mean and other quantities that appear
in standard RANS models, ensuring that there is no confounding between
typical RANS closures and additional closures required for the slow
growth problem.  The performance of the method is demonstrated on two
problems: an essentially incompressible, zero-pressure-gradient
boundary layer and a transonic boundary layer over a cooled wall with
wall transpiration.  The results show that the approach produces flows
that are qualitatively similar to other slow growth methods as well as
spatially developing simulations and that the new method can be a
useful tool in investigating complex wall--bounded flows.

\commentout{
A slow growth formulation for DNS of temporally evolving boundary layers is
presented and demonstrated. 
The formulation relies on a multiscale approach to account separately for the
slow time evolution of statistical averages, and the fast time evolution of
turbulent fluctuations.
The source terms that arise from the multiscale analysis are modeled assuming a
self-similar evolution of the averages.
The performance of the formulation is evaluated using DNS of spatially evolving
compressible boundary layers. 
This formulation was developed to provide data for the calibration of turbulence
model parameters and enable the quantification of uncertainty due to the models. 
\todo[inline]{Review, make more specific to final content of paper \ldots}
}
\end{abstract}

\maketitle

\section{Introduction} \label{sec:introduction}
Direct numerical simulation (DNS) is a valuable tool for investigating
turbulent boundary layers. DNS is of particular value to the
formulation, calibration, and testing of engineering turbulence
models, such as Reynolds Averaged Navier-Stokes (RANS) models, because
the conditions in which the turbulence evolves are precisely defined,
making it possible for model-based simulations to be performed under conditions that
exactly match those in which the data is generated. Another important
use of boundary layer DNS is the study of the structure and statistics
of the turbulence. In this case, the ability to access three
dimensional time-dependent turbulent velocity and scalar fields is
of great value. Furthermore, experimental measurements in turbulent
boundary layers are often difficult and limited, especially in the
presence of complicating features such as transpiration,
compressibility, and chemical reactions. In these situations, DNS can
provide data that would not otherwise be available.  In this work, we
aim to develop DNS model problems that are 1) well-suited to
generating data for turbulence model calibration and testing in
boundary layers and 2) easily generalizable to complex situations to
enable the study of complex boundary layers.

\commentout{
One of our primary concerns in this paper
is the development of DNS model problems that are well suited to
generating data for turbulence model calibration and testing in
boundary layers.

 So our second concern in this paper, is the development of
DNS model problems that are a good vehicle for the study of complex
boundary layers.
}

The DNS of spatially developing boundary layers, which most often
occur in reality, presents challenges. The biggest issue is the very
long evolution lengths that are required for the turbulence to
equilibrate and eliminate artifacts of artificial inlet boundary
conditions.  This issue also arises in experiments, where a long
distance is required for a boundary layer to relax to a canonical
turbulent boundary layer downstream of a trip.  However, in the case
of DNS, this long evolution requires very large computational domains
and, consequently, great computational costs \cite{Sillero2013}. The importance of this
issue was highlighted by~\citet{schlatter2010assessment} who found
that, even considering only well-resolved simulations, results for DNS
of incompressible, low Reynolds number, turbulent boundary layers show
disconcerting inconsistencies.  They concluded that the discrepancies
are due to difficulties associated with spatially developing
simulations, including limited domain sizes and inflow boundary data.

The required streamwise domain size of a DNS of a spatially evolving
boundary layer can be minimized with realistic inflow boundary
conditions.  Formulation of appropriate inflow conditions for
spatially evolving simulations is a well-known problem.  Often, an
auxiliary simulation or a recycling/rescaling procedure is used.
While such procedures have been the subject of ongoing research for
over 20 years~\citep{Wu2017}, they still introduce implementation
complexities and modeling challenges.  For instance, even in the best
understood scenario, a canonical zero-pressure-gradient flat plate boundary
layer, where the method of~\citet{Lund1998} has been used
successfully, recycling/rescaling procedures have the potential to
introduce spurious periodicity~\citep{Nikitin2007} and other
issues~\citep{Jewkes2011}.  In cases with additional complicating
phenomena, such as wall transpiration or chemical reactions, the
challenges associated with posing appropriate inflow conditions can
only increase.

Motivated by the difficulties of simulating spatially evolving
boundary layers,~\citet{spalart1988direct} developed a ``slow growth''
approximation, in which the effects of the slow streamwise evolution are
modeled while the turbulent fluctuations are directly simulated.  Slow growth
approaches rely on an assumed separation of
scales between the fast evolution of the turbulent fluctuations and
the slow evolution of mean characteristics of the boundary layer. Because of this
separation, one can conduct a DNS of the fast evolution at
a single, fixed point in the slow evolution, with slow evolution
effects modeled.
%
In a slow growth formulation, the fast scale turbulence becomes
homogeneous in the streamwise direction. This allows the use of
periodic boundary conditions in the streamwise direction, eliminating
the need for turbulent inflow boundary conditions or an exceptionally
long streamwise domain size. Further, homogeneity enables the use of
Fourier spectral methods, which are the preferred numerical
discretizations for DNS due to their efficiency and good resolution
properties.

In the work presented here, a new slow growth DNS model is developed
and applied.  The approach is based on homogenization in time, rather
than space, and an assumption of self-similarity in the slow
evolution.  It is constructed to support calibration and validation of
RANS turbulence models for compressible boundary layers with
transpiration and to be generalizable to boundary layers with
complicating physical phenomena such as chemical reactions and
favorable pressure gradients.  Naturally, because of the
approximations required to formulate a slow growth DNS, the resulting
homogenized boundary layer will necessarily differ from a spatially
developing layer.  As will be shown in Section~\ref{sec:results}, the
temporal slow growth turbulent boundary layers obtained here resemble
spatially evolving layers to a degree comparable to previous spatially
homogenized boundary layers.  Whether the remaining differences are
important depends on the goals of the simulation.  If the goal is to
learn as much as possible about the features of a particular spatially
evolving flow, then a spatially developing simulation is best.  In
this case, it is worth the time and effort required to overcome the
challenges associated with inflow boundary conditions and long domain
sizes noted previously.


\commentout{
\todo{This feels repetitive.  Can we fold in with above?}
The slow growth idea relies on the assumption that there exists a
separation of scales between the fast evolution of the turbulent
fluctuations and the slow evolution of mean flow quantities. Because
of this separation, one can imagine conducting a DNS of the fast
evolution at a single, fixed point in the slow evolution.  Such a
simulation requires that the effects of the slow evolution be modeled.
In developing such a model, it is convenient to assume that the slow
evolution occurs in a self--similar fashion.
}


However, if the goal is to learn more generally about features of
wall--bounded turbulence---including, for example, the ability of RANS
models to represent the effects of turbulence in such flows or how the
turbulence is affected by complicating physical phenomena (e.g.,
chemistry)---it is not necessarily crucial to simulate a spatially
developing boundary layer.  Instead, there are two requirements.
First, the fast turbulent scales must be governed by the
Navier--Stokes equations with forcing provided by the slow evolution.
Second, the modeled effect of the slow evolution must be sufficiently
representative of the flow of interest.  Thus, it is not necessary
that the effects of the slow evolution be represented exactly, and in
fact, one may be willing to tolerate differences in the name of
simplicity if their effects can be understood.  This realization
enables the development of a slow growth modeling approach that is
easily extensible to increasingly complex physical phenomena, allowing
straightforward and computationally efficient investigations of the
effects of these complicating phenomena on wall-bounded turbulence.

\subsection{Previous Slow Growth Formulations} \label{sec:previous}
By modeling the forcing due to the slow evolution, slow growth
homogenization formulations enable efficient simulation of turbulence that is
representative of that in an evolving flow.  The
slow growth simulation concept was pioneered for incompressible turbulent
boundary layers in a series of papers~\citep{Spalart1985,
  Spalart1986} which culminated in simulation of an incompressible,
zero-pressure-gradient turbulent boundary layer with Reynolds number
up to $Re_{\theta} =
1410$~\citep{spalart1988direct}.  The approach was later extended to
compressible flows by~\citet{guarini2000direct} and used to simulate a
$M_{\infty} = 2.5$, $Re_{\theta} = 1577$, adiabatic wall boundary
layer.  Both~\citet{spalart1988direct} and~\citet{guarini2000direct}
formulated slow growth models based on a coordinate transform combined
with a multi-scale analysis.  In these approaches, the coordinate
transformation is designed to fit the boundary layer growth, with the goal
that, for a section of small streamwise extent, the flow is
approximately homogeneous in the transformed streamwise direction.  Then, a
multi-scale analysis is performed to split the streamwise variation
into slow and fast components.  The result of the analysis is a set of
equations governing the fast component of the flow at a single point
in the slow streamwise evolution.  These equations are formally
equivalent to the Navier--Stokes equations with the addition of source
terms that quantify the effect of the slow evolution.  Then, to enable
a slow growth simulation, the source terms are modeled to close the
system.

In the context of the current work, the existing slow growth formulations have two
main drawbacks.  First, as formulated
by~\citet{guarini2000direct}, many modeling assumptions are required
in the compressible regime.  For instance, the van Driest relationship
is used to relate mean temperature and streamwise velocity, and it is
assumed that the van Driest transformed velocity satisfies typical
incompressible scaling laws.  These assumptions do not necessarily
hold for more general situations, and it is unclear how to extend
the formulation to such cases.

The second difficulty is specific to using the data
resulting from slow growth DNS for calibration and validation of RANS
turbulence models.  In doing so, one will naturally be required to
solve the Reynolds--averaged slow growth equations, which are obtained
by applying the Reynolds averaging procedure to the slow growth
equations.  The resulting equations govern the mean flow at a
particular point in the slow evolution and contain all the usual
unclosed terms---e.g., the Reynolds stress---as well as the Reynolds
average of the slow growth sources, which represent the mean forcing
provided by the slow evolution.  Thus, to avoid confounding errors
introduced by the standard RANS closures with those introduced by
additional models required to close the mean slow growth sources, it
is necessary for the mean slow growth source terms to be closed purely
in terms of the mean flow and quantities that are already modeled as
part of a standard RANS model.  Neither the Spalart nor the Guarini
formulations satisfy this requirement.

\subsection{Overview}
To overcome these limitations of existing slow growth DNS models, a
new formulation is developed and presented in this work.  The approach
is based on homogenization of a temporally evolving boundary layer.
Thus, the motivating flow is the classical temporal boundary layer,
where an infinite plate is impulsively started at time $t=0$.  In this
situation, a boundary layer develops over the plate.  This boundary
layer is naturally homogeneous in the streamwise and spanwise
directions, inhomogeneous in the wall--normal direction, and
non--stationary since it grows in time.  Thus, unlike the approaches
of~\citet{spalart1988direct} and~\citet{guarini2000direct}, this
formulation requires homogenization in time rather than space.  This
switch enables the development of more easily extensible models for
the slow growth forcing terms.  Section~\ref{sec:formulation} gives
details of this formulation, including constraints imposed by the RANS
calibration and validation use case and the specific modeling
assumptions invoked to develop a concrete model.  Then, two sets of
example results are reported in Section~\ref{sec:results}.  To show
how the results of the present formulation differ from previous slow
growth models, Section~\ref{sec:results_lowM} compares statistics from
the present formulation for a $M_{\infty} = 0.3$ turbulent boundary
layer to those from a slow growth simulation due to
\citet{spalart1988direct} and a spatially evolving simulation due to
\citet{schlatter2010assessment}.
To
demonstrate the applicability of the approach to more complex flows,
Section~\ref{sec:results_highM} shows statistics from a transonic
turbulent boundary
layer with a cold wall and wall transpiration.  The cold wall and
transpiration are seen to have dramatic effects on both the mean
velocity profile and turbulence quantities near the wall.
Section~\ref{sec:conclusions} provides conclusions and directions for
future work.


\section{Temporal Slow Growth Formulation} \label{sec:formulation}
This section describes a temporal slow growth DNS model designed to
yield data useful for calibration and validation of RANS models. In
the development to follow, $\rho$ will denote the fluid density, $u_i$
the velocity vector in Cartesian tensor notation, and $E=e+u_ku_k/2$ the total
energy per unit mass, including the internal energy ($e$) and the
kinetic energy. Einstein summation convention will be used throughout.
The spatial position vector is $x_i$, with the wall-normal coordinate
also designated as $y$. Reynolds averaging will be denoted by an
overbar, and the Reynolds fluctuations by a single prime. Thus, the
Reynolds decomposition of the density is given by
$\rho=\overline{\rho}+\rho'$. The Favre, or density-weighted, average
will be denoted by a tilde, and the Favre fluctuations by a double
prime. So, the Favre decomposition of the velocity is given by
$u_i=\tilde u_i+u''_i=\overline{\rho u}_i/\overline{\rho}+u''_i$.


\subsection{Multi-scale Formulation and RANS} \label{sec:overview}
As described in Section~\ref{sec:introduction}, a statistically
stationary slow growth model is sought for a
temporally evolving turbulent boundary layer developing over an impulsively
started infinite flat plate.  The evolution of such a
boundary layer is described by the compressible Navier--Stokes
equations, written here in a generic form that will facilitate
the analysis to follow:
\begin{equation}
\pp{\rho q}{t} + \mathcal{N}_{\rho q} = 0.
\end{equation}
Here, $q$ represents one of the five conserved quantities per unit
mass. That is, $q$ is either 1, one of the velocity components $u_i$ or
the total energy per unit mass $E$, so that the volume density of the
conserved quantities are $\rho$ for mass, $\rho u_i$ for momentum, and
$\rho E$ for energy. The quantities $\rho$ and $q$ make up the
so-called primitive variables. The symbol $\mathcal{N}_{\rho q}$ then
represents all the remaining terms in the equation for $\rho q$ in the
Navier-Stokes equations. For example, $\mathcal{N}_\rho=\partial \rho
u_i/\partial x_i$.

The slow growth formulation developed here is based on the assumption
that the boundary layer grows much more slowly than the evolution of
the turbulence. This motivates the use of a multi-time-scale asymptotic
formulation in terms of a fast time $t_f=t$ and a slow time
$t_s=\epsilon t$, where $\epsilon\ll 1$. The turbulence fluctuations
are presumed to evolve in fast time $t_f$, whereas mean quantities
evolve only in slow time $t_s$. Introducing this two-time formulation
into the Navier-Stokes equations yields
\begin{equation}
\label{eqn:slowgrowthgeneric}
\pp{\rho q}{t_f} + \mathcal{N}_{\rho q} = - \epsilon \pp{\rho q}{t_s}. 
\end{equation}
The objective is to perform a DNS of the Navier-Stokes equations in
fast time $t_f$ at some constant value of the slow time $t_s=\tz$. For
an impulsively started plate, the boundary layer thickness is just a
function of $t_s$, and so specifying $t_s=\tz$ is equivalent to
defining the boundary layer thickness and therefore the Reynolds
number of the DNS. The DNS will thus solve the equations
\begin{equation}
\pp{\rho q}{t_f} + \mathcal{N}_{\rho q} 
= \Ssd_{\rho q} =  \left. - \epsilon \pp{\rho q}{t_s}\right|_{t_s=\tz},
\label{eqn:fast_time_evolve}
\end{equation}
where $\Ssd_{\rho q}$ is referred to as the slow growth source, which
must be modeled. In addition to \eqref{eqn:fast_time_evolve}, it will
be convenient to consider the primitive variable form of the slow growth
Navier-Stokes equations
\begin{gather}
\pp{\rho}{t_f} + \mathcal{N}_{\rho} = \Ssd_{\rho} , \label{eqn:slow_growth_rho} \\
\pp{q}{t_f}    + \mathcal{N}_{q}    = \Ssd_{q} \label{eqn:slow_growth_q}.
\end{gather}
where in the usual way
\begin{gather}
  \mathcal{N}_q=\frac{1}{\rho}(\mathcal{N}_{\rho
    q}-q\mathcal{N}_\rho),\\
  \Ssd_q=\frac{1}{\rho}(\Ssd_{\rho q}-q\Ssd_\rho).\label{eq:RANSSource}
\end{gather}


%

In formulating models for the slow growth source $\Ssd_{\rho q}$, it
will be important to consider how the source terms enter the RANS
equations. If the sources in the RANS equations are closed with
respect to the RANS state variables, then a RANS of the resulting slow
growth system will not require any additional modeling assumptions
besides those inherent to the RANS model. The RANS equations are
obtained by averaging the Navier-Stokes equations.  Because the
temporally homogenized turbulent boundary layer will be statistically
stationary, this procedure gives simply
\begin{equation}\label{eq:GenericRANS}
  \mean{\mathcal{N}_{\rho q}} = \mean{\Ssd_{\rho q}}.
\end{equation}
In addition, RANS models generally involve one or more auxiliary
equations for turbulence quantities, such as the turbulent kinetic
energy per unit mass $k=\widetilde{u''_iu''_i}/2$ and the turbulent energy dissipation rate per
unit mass $\epsilon$, in the $k$-$\epsilon$ equation. Another common
auxiliary equation in RANS models is the equation for the Reynolds
stress tensor $R_{ij}=\mean{\rho u''_iu''_j}$. Because,
$k=R_{ii}/2\mean{\rho}$, it will be sufficient to consider just the
Reynolds stress equation, which reduces to
\begin{equation}
\label{eq:Reynolds_stresses}
\underbrace{\overline{\ffluc{u_i} \ffluc{u_j} \mathcal{N}_{\rho} + \rho \ffluc{u_i}
\mathcal{N}_{u_j}     + \rho \ffluc{u_j} \mathcal{N}_{u_i}
}}_{\overline{\mathcal{N}_{R_{ij}} }}
= \underbrace{\overline{\ffluc{u_i} \ffluc{u_j} \Ssd_{\rho} + \rho \ffluc{u_i} \Ssd_{u_j} + \rho \ffluc{u_j} \Ssd_{u_i} }}_{\overline{\Ssd_{R_{ij}}}} .
\end{equation}
To avoid RANS modeling of terms arising from the slow growth source
terms, we will require that the right hand sides of
(\ref{eq:GenericRANS}-\ref{eq:Reynolds_stresses}) be closed in terms
of the RANS state variables.

For simplicity, we do not require that the slow growth source term in the dissipation rate equation
be closed.  However, for constant density,
constant viscosity flows, the formulation shown in
Section~\ref{sec:construct_model} does result in
a dissipation equation slow growth source
that is closed in terms of $\epsilon$ and $k$.  This result does not
hold for a general compressible flow.  However, in non-hypersonic wall-bounded flows, the
dissipation is dominated by the solenoidal component~\cite{Huang1995,
  guarini2000direct, Candler2003}. We therefore expect that a closure model
based on the incompressible result would adequately model the effect
of the slow growth sources for many cases of interest.

\subsection{RANS-Consistent Slow Growth Sources} \label{sec:consistent}
As is shown in Appendix~\ref{sec:inconsistent}, a straightforward
formulation of the slow growth sources in terms of the conserved
variables leads to sources in the RANS equations that are
unclosed. Here it is shown that a formulation based on the primitive
variable source terms can yield RANS source terms that are closed.
Consider the following slow growth source term formulation:
\begin{align}
\label{eq:Srho_consistent_form}
\Ssd_{\rho} &= \rho f_{\rho}, \\
\label{eq:Sq_consistent_form}
\Ssd_{q}    &= g_{q} + \ffluc{q} h_{q},
\end{align}
where the functions $f_\rho$, $g_q$, and $h_q$ depend only on $y$ and
are expressed in terms of the statistical quantities that serve as
state variables in the  RANS
models. When these forms are used to write the RANS slow growth sources
in (\ref{eq:GenericRANS}) using (\ref{eq:RANSSource}), the results are
\begin{gather*}
\mean{\Ssd_{\rho}} = \mean{\rho} f_\rho, \\
\mean{\Ssd_{\rho q}} 
= \mean{q \Ssd_{\rho}}    + \mean{\rho  \Ssd_{q}}
= \mean{\rho q}  f_{\rho} + \mean{\rho} g_{q} + \underbrace{\mean{\rho \ffluc{q}}}_{=0}  h_{q}
= \mean{\rho q}  f_{\rho} + \mean{\rho} g_{q}.
\end{gather*}
Thus, the mean slow growth sources are closed purely in terms of the
RANS variables $\mean{\rho}$, $\mean{\rho q}$, and the dependencies of
$f_{\rho}$ and $g_q$. Similarly expanding the source in the Reynolds
stress transport equations---i.e., the right hand side
of~\eqref{eq:Reynolds_stresses}---yields
\begin{align}
\label{eq:S_Reynolds_stresses}
\mean{\Ssd_{R_{ij}}} 
&= \mean{\ffluc{u_i} \ffluc{u_j} \Ssd_{\rho}} 
 + \mean{\rho \ffluc{u_i} \Ssd_{u_j}}
 + \mean{\rho \ffluc{u_j} \Ssd_{u_i}} \notag \\
&= \mean{R_{ij}} f_{\rho}
 + \underbrace{\mean{\rho \ffluc{u_i}} g_{u_j}}_{=0} + \mean{R_{ij}} h_{u_j}
 + \underbrace{\mean{\rho \ffluc{u_j}} g_{u_i}}_{=0} + \mean{R_{ij}} h_{u_i}.
\end{align}
In this case, the Reynolds stress slow growth source is closed only in
terms of the Reynolds stress tensor, and the dependencies of $f_\rho$
and $h_{u_i}$.

Note also that $\mean{\Ssd_{R_{ij}}}$ is a second-rank tensor, and so
the right hand side of (\ref{eq:S_Reynolds_stresses}) must be as
well. This can only be true if the function $h_{u_i}$ is a scalar,
that is, it is the same function $h_u$ for all $i$. Similarly
considering that $\mean{\Ssd_{\rho u_i}}$ is a vector, it is clear
that $g_{u_i}$ must be a vector.  These conditions that lead to
tensorial consistency will be used in choosing the final form of the
models in Section~\ref{sec:construct_model}.

As discussed in Section~\ref{sec:overview}, RANS models often carry
equations for the turbulent kinetic energy $k$ rather than the
Reynolds stress tensor. Since the closure of
$\mean{\Ssd_{R_{ij}}}$ is in terms of the Reynolds
stress tensor, and $k=R_{ii}/2\rho$, the slow growth source in the
turbulent kinetic energy equation
$\mean\Ssd_{\rho k}=\Ssd_{R_{ii}}/2$ will be closed in
terms of $k$, provided $h_u$ depends on $R_{ij}$ only through $k$.

%
%


\subsection{Constructing the slow growth model} \label{sec:construct_model}
The development in Section~\ref{sec:consistent} shows how the slow
growth source model can yield RANS sources that are closed.  However,
it does not determine an actual model.  In this section, a model of
the form shown in
(\ref{eq:Srho_consistent_form}-\ref{eq:Sq_consistent_form}) is
developed based on a multi-time-scale expansion of the primitive
variables, analogous to the spatial expansion introduced
by~\citet{spalart1988direct} and~\citet{guarini2000direct}.
  
%
The multi-scale expansions of $\rho$ and $q$ are formulated in terms of the mean and
fluctuations as follows:
\begin{align}
\label{eqn:rho_decomposition}
\func{\rho}{x, y, z, t} = 
  \func{\mean{\rho}}{y, t_s} +
  \underbrace{ 
   \func{A_{\rho}}{y,t_s} \func{\fluc{\rho}_p}{x,y,z,t_f}
  }_{\func{\fluc{\rho}}{x,y,z,t_f,t_s}},\\
\label{eqn:flow_decomposition}
\func{q}{x, y, z, t} = 
  \func{\fav{q}}{y, t_s} +
  \underbrace{ 
   \func{A_{q}}{y,t_s} \func{\ffluc{q}_p}{x,y,z,t_f}
  }_{\func{\ffluc{q}}{x,y,z,t_f,t_s}}.
\end{align}%
%
Here, $A_\rho$ and $A_{q}$ are amplitude functions which characterize the
magnitude of the fluctuations. Also,  $\fluc{\rho}_p$ and $\ffluc{q}_p$ are the
turbulent fluctuations normalized by this amplitude.
Consistent with the association of the slow time with growth of the
boundary layer in time, 
we assume that $\overline{\rho}$, $\fav{q}$, $A_\rho$ and $A_q$ vary
only on the slow time scale, while
$\rho'$ and $\ffluc{q}_p$ vary on the fast time scale.

Using (\ref{eqn:flow_decomposition}), the time derivative of $q$ can be expressed as
\begin{equation*}
%
\pp{q}{t} 
= \pp{q}{t_f} + \epsilon \left(\pp{\fav{q}}{t_s} + \pp{\ffluc{q}}{t_s}\right)
= \pp{q}{t_f} + \epsilon \left(\pp{\fav{q}}{t_s} + \frac{\ffluc{q}}{A_q} \pp{A_q}{t_s} \right).
\end{equation*}
From this result, it is clear that the slow growth source term $\Ssd_q$ is
simply
\begin{equation}\label{eq:SqModel1}
\Ssd_q=-\epsilon \left(\pp{\fav{q}}{t_s} + \frac{\ffluc{q}}{A_q}\pp{A_q}{t_s} \right).
\end{equation}
The challenge then is to model the slow time derivatives of $\fav{q}$
and $A_q$. To do so, we assume that $\fav{q}$ and $A_q$ evolve
self-similarly in slow time; that is:
\begin{align}
\label{eqn:similarMean}
\func{\fav{q}}{t_s, y} &= \func{F_q}{\frac{y}{\Delta(t_s)}}, \\
\label{eqn:similarRms}
\func{A_q}{t_s, y}     &= \func{G_q}{\frac{y}{\Delta(t_s)}},
\end{align}
where $\Delta(t_s)$ is a measure of boundary layer thickness. Note
that this self-similar form is not exactly satisfied by a
time-evolving turbulent boundary layer because, as is well known, the
thickness of the near-wall layer grows much more slowly than the
overall boundary layer thickness. Further, the magnitude of the
turbulent fluctuations also evolve with the growth of the layer,
albeit slowly. Despite these shortcomings, the above similarity forms
will temporally homogenize the turbulent boundary layer, and produce a
flow with many of the characteristics of an evolving boundary layer,
as is shown in Section~\ref{sec:results}. Also, the DNS model
developed from these assumptions will be closed given typical RANS
variables, and thus will meet the goal of supporting RANS model
development.

Introducing the similarity forms (\ref{eqn:similarMean}-\ref{eqn:similarRms})
into the right hand side of (\ref{eq:SqModel1}) yields
\begin{equation}\label{eq:SqModel2}
\Ssd_q=y \left(\frac{\epsilon}{\Delta} \pp{\Delta}{t_s}\right) 
             \pp{\fav{q}}{y}+\ffluc{q}
      y \left(\frac{\epsilon}{\Delta} \pp{\Delta}{t_s}\right) 
              \frac{1}{A_q} \pp{A_q}{y},
\end{equation}
The logarithmic derivative of $\Delta$ that appears in the parentheses
is just the exponential growth rate of the boundary layer, which is a
function of time. Or, because $\Delta$ is a monotonically increasing
function of time, the growth rate $\grt$ can be considered a function of
$\Delta$:
\begin{equation}
\grt(\Delta)=\left( \frac{1}{\Delta} \pp{\Delta}{t} \right),
\end{equation}
where the slow time derivative has been expressed in terms of the
physical time derivative, using the fact that $\Delta$ varies only in
slow time. In a slow growth homogenized DNS, the boundary layer thickness will
remain constant, so that $\grt$ will also be a constant. Indeed, it
is the only parameter that needs to be specified in the slow growth
source model. Once one determines the desired Reynolds number and
therefore the boundary layer thickness $\Delta$, the function
$\grt(\Delta)$ determines the required value of the constant. However,
the function is not known \emph{a priori}, so in practice, we commonly
use an auxiliary RANS computation to determine a value of $\grt$
that will yield a value of $\Delta$ close to that specified.

Comparing (\ref{eq:SqModel2}) to (\ref{eq:Sq_consistent_form}), it is clear
that the two forms are consistent, provided that the functions $g_q$
and $h_q$ are given by
\begin{gather*}
g_q = y \, \grt \pp{\fav{q}}{y},\\
h_q = y \, \grt \frac{1}{A_q} \pp{A_q}{y}.
\end{gather*}
Therefore, provided the $A_q$ are defined in terms of RANS state
variables, and the tensor consistency conditions are met, the source
model will result in consistent closed source terms in the RANS equations.
To meet these requirements, and in recognition of the fact that the
root-mean-square (RMS) of the fluctuation velocity and total energy measure
the strength of the fluctuations, $A_{u_i}$ is taken to be the same scalar
$A_u$ for all values of $i$,
\begin{gather}\label{eq:Au}
A_u=\sqrt{\fav{\ffluc{u_k}\ffluc{u_k}}}=\sqrt{2k},
\end{gather}
and,
\begin{gather}
\label{eq:AE}
A_E=\sqrt{\fav{\ffluc{E} \ffluc{E}}}.
\end{gather}
The resulting dependence of $h_u$ on $k$ is exactly what was required
to ensure closure of the source term in the Reynolds stress transport
and turbulent kinetic energy equations. There is no such restriction
on $A_E$, since this term does not contribute to the mean of $\Ssd_E$.
Finally, note that because $\fav{q}$, $\sqrt{2k}$ and
$\sqrt{\fav{\ffluc{E} \ffluc{E}}}$ are fields with no variation in
the directions parallel to the wall, the 
$y\frac{\partial}{\partial y}$ operators in (\ref{eq:Au}-\ref{eq:AE})
can be written as $x_i\partial/\partial x_i$, the inner product of the
coordinate vector ${\bf x}$ with the gradient operator. This makes
clear that $g_{u_i}$ is a vector, and $h_u$, $g_E$ and $h_E$ are
scalars, as required for tensor consistency.

To complete the model, it remains to construct the slow growth source
for conservation of mass.  Following similar steps beginning
from~\eqref{eqn:rho_decomposition}, we have
\begin{align}
\Ssd_{\rho} 
    &= y \, \grt \pp{\mean{\rho}}{y}
     + \fluc{\rho} y \, \grt \frac{1}{A_\rho} \pp{A_\rho}{y}.
\label{eq:rho_source_general}
\end{align}
Then, choosing $A_\rho = \mean{\rho}$, the source for density is
consistent with the form~\eqref{eq:Srho_consistent_form}.
\begin{equation*}
\Ssd_{\rho} 
= \mean{\rho} y \, \grt \frac{1}{\mean{\rho}} \pp{\mean{\rho}}{y} + \fluc{\rho} y \, \grt \frac{1}{\mean{\rho}} \pp{\mean{\rho}}{y}
= \rho \underbrace{y \, \grt \frac{1}{\mean{\rho}} \pp{\mean{\rho}}{y}}_{f_\rho}.
\end{equation*}
%

\subsection{Summary of equations}
In summary, the complete set of slow growth Navier--Stokes equations
used in
this work are given by,
\begin{align}
\label{eqn:ns_homogenized_rho}
\pp{\rho}{t_f} + \pp{}{x_i} (\rho u_i) 
   &= \Ssd_\rho, \\
\label{eqn:ns_homogenized_rui}
\pp{}{t_f}(\rho u_i) + \pp{}{x_j}(\rho u_j u_i) 
   &= - \pp{p}{x_i} + \pp{\tau_{ji}}{x_j} 
      + \rho \Ssd_{u_i} + u_i \Ssd_{\rho} , \\
\label{eqn:ns_homogenized_rE}
\pp{}{t_f}(\rho E) + \pp{}{x_j}(\rho u_j H) 
   &= \pp{}{x_j}(\tau_{ji} u_i) - \pp{q_j}{x_j} 
      + \rho \Ssd_{E} + E \Ssd_{\rho} ,
\end{align}
where $p$ is the
pressure, $\tau_{ij}$ is the viscous stress tensor, $q_j$ is the heat flux
vector, and $H = h + u_k u_k / 2$ is the total enthalpy per unit mass, with $h$ the
enthalpy per unit mass. 
The slow growth sources are modeled as
\begin{align}
\label{eqn:slowGrowthFinal}
\Ssd_{\rho} 
    &= \rho y \, \grt 
       \frac{1}{\mean{\rho}} \pp{\mean{\rho}}{y}, \\
\Ssd_{u_i} 
    &= y \,\grt
          \left(
             \pp{\fav{u_i}}{y} 
             + \frac{\ffluc{u_i}}{\sqrt{\fav{\ffluc{u_k} \ffluc{u_k}}}} \pp{\sqrt{\fav{\ffluc{u_k} \ffluc{u_k}}}}{y}
          \right), \\
\Ssd_{E} 
    &= y \, \grt
          \left(
             \pp{\fav{E}}{y} 
             + \frac{\ffluc{E}}{\sqrt{\fav{\ffluc{E} \ffluc{E}}}} \pp{\sqrt{\fav{\ffluc{E} \ffluc{E}}}}{y}
          \right).
\end{align}
When coupled with appropriate models for the thermodynamics (e.g.,
ideal gas) and viscous transport (e.g., Newtonian fluid with
Sutherland's law), these equations constitute a closed system that
allows one to perform DNS using the temporal slow growth formulation.

%


\section{Results} \label{sec:results}

To illustrate the temporal slow growth DNS model described in
Section~\ref{sec:formulation}, results for two cases are presented.
The first case, reported in Section~\ref{sec:results_lowM} and denoted
Case (\lowM), is a low Mach number ($M_\infty = 0.3$, essentially
incompressible) boundary layer to enable comparison with the spatially
homogenized boundary layer of~\citet{spalart1988direct} and the
spatially evolving simulation reported
by~\citet{schlatter2010assessment}.  The second case, reported in
Section~\ref{sec:results_highM} and denoted Case (\cev), is a
transonic boundary layer with a strongly cooled, blowing wall.  The
conditions for this case---namely the edge Mach number $M_{\infty} = 1.2$,
the ratio of the wall temperature to the adiabatic wall temperature of
$T_{\wall} / T_{\awall} = 0.23$, and the blowing velocity normalized
by the friction velocity of $v^+_{\wall} = 0.0188$---were inspired by
features of the boundary layer that develops on a space capsule with
an ablating thermal protection system during atmospheric
entry~\citep{Bauman2011,Stogner2011,Kirk2014}.  The case demonstrates
the ease with which complications from a highly cooled, blowing wall
can be incorporated into the temporal slow growth formulation.

For both cases, the working fluid is taken to be calorically perfect
air, and the viscosity is computed according to Sutherland's
law~\citep{White_1991_Viscous_Flow}: $\mu = C_1 T^{3/2} / (T + S)$
where $C_1 = \mu_0 T_0^{-3/2}
(T_0+S)$=\num{1.458e-6}\si{\pascal-\second\per\kelvin^{0.5}} and
$S$=110.4\si{\kelvin}.  Further details of the case scenarios and
grids are given in
Tables~\ref{tab:ScenarioParameters},~\ref{tab:DNSParameters},
and~\ref{tab:BLParameters}.  In the tables and throughout the
discussion to follow, $M$ denotes Mach number; $Re$ denotes Reynolds
number; $T$ is temperature; $v$ is the wall-normal velocity component.
The subscript $()_\wall$ denotes wall conditions; the subscript
$()_{\infty}$ denotes freestream conditions; and the superscript
$()^+$ denotes non-dimensionalization by the usual viscous scales
(i.e., the friction velocity $u_{\tau} = \sqrt{\tau_\wall /
  \rho_\wall}$, where $\tau_\wall$ is the shear stress at the wall,
and the kinematic viscosity at the wall, $\nu_\wall$).  Boundary layer
length scales are denoted by $\theta$ for the momentum thickness,
$\delta^*$ for the displacement thickness; and $\delta$ for the
distance from the wall where the streamwise mean velocity obtains 99\%
of the freestream value.  $H_1 = \delta^*/\theta$ is the shape factor,
$H_2 = \delta / \theta$, and $c_f = 2 \tau_\wall /(\rho_{\infty}
u_{\infty}^2)$ is the skin friction coefficient.  The domain size is
denoted by $L_x$, $L_y$, and $L_z$ in the streamwise, wall-normal, and
spanwise direction, respectively, and the total number of points in
each direction is denoted by $N_x$, $N_y$, and $N_z$.  The distance from the wall to 
the first grid point is $y_1$, and $N_{y<y_{10}^+}$ and
$N_{y<\delta}$ are the number of wall-normal points inside $y^+ = 10$
and $y=\delta$.



%
\begin{table}
\caption{\small{Flow and scenario parameters for the temporal slow growth DNS cases.}}
\centering
\begin{tabular}{lcccccccc}
\hline
  Case          & M$_\infty$ & Re$_\theta$ &  Re$_{\tau}$ & $T_\wall/T_\awall$
& $v_\wall^+$ & $T_\wall [\si{\kelvin}]$  & $T_\infty [\si{\kelvin}]$ & $\grt(\Delta) [\si{\per\second}]$ \\
\hline
  (\lowM) &  0.3       &  703        &   306      &   1.0  &  
  0.0        & 5500 &  5500    & 65 \\
  (\cev)  &  1.2       &  422        &   685      &   0.23 &  
  0.0188     & 1634 &  5604   & 330  \\
\hline
\end{tabular}
\label{tab:ScenarioParameters}
\end{table}
\begin{table}
\caption{\small{Domain size and grid parameters for the temporal slow growth DNS cases.}}
\centering
\begin{tabular}{lccccccc}
\hline
  Case    
& $L_x/\delta \times L_y/\delta \times L_z/\delta$
& $N_x \times N_y \times N_z $
& $\Delta_x^+$
& $\Delta_z^+$
& $y_1^+$
& $N_{y<y_{10}^+}$
& $N_{y<\delta}$
\\ 
\hline                                                                                   
  (\lowM) &  $\num{11.7 x 2.9 x 3.5}$ & $\num{256 x 205 x 128}$ & 14.01
&   8.43  &  0.61 & 17 & 129 \\
  (\cev)  &  $\num{10.6 x 2.6 x 3.2}$ & $\num{448 x 370 x 256}$ & 16.14  
&   8.53  &  0.63 & 17 & 246 \\
\hline
\end{tabular}
\label{tab:DNSParameters}
\end{table}
\begin{table}
\caption{\small{Boundary layer parameters for the temporal slow growth DNS cases.}}
\centering
\begin{tabular}{lccccccc}
\hline
  Case    & Re$_\theta$ &  Re$_\delta^*$ & Re$_{\tau}$   & H$_1$ &  H$_2$ & $c_f$         \\ 
\hline
  (\lowM) &  703        &   1050         &  306          & 1.49  &  8.98  & \num{4.70e-3} \\
  (\cev)  &  422        &   267          &  685          & 0.63  &  7.22  & \num{4.65e-3} \\
\hline
\end{tabular}
\label{tab:BLParameters}
\end{table}

Both simulations were performed using the compressible DNS code
Suzerain~\citep{Ulerich2014}.  The spatial discretization in Suzerain
couples a Fourier/Galerkin discretization in the periodic streamwise
and spanwise directions with a B-spline/collocation method in the
wall-normal direction.  The time advance is accomplished using a
semi-implicit Runge-Kutta scheme in which only the mean wall-normal
convective and viscous terms are treated implicitly.
See~\citet{Ulerich2014} for further details regarding numerical
methods and the code.

\subsection{Case (\lowM): $M_{\infty}=0.3$ Boundary Layer} \label{sec:results_lowM}
Statistics from the Case (\lowM) simulation are presented in this section.
For comparison, the temporal slow growth boundary layer at
Re$_\theta = 703$ is compared with the spatial slow growth case at
Re$_\theta = 670$ from \citet{spalart1988direct} and a spatially evolving
boundary layer by \citet{schlatter2010assessment} at Re$_\theta = 677$.  At
this condition, the temporal slow growth DNS produces global
boundary layer metrics similar to those reported for the
spatial slow growth and spatially developing simulations, as shown in
Table~\ref{tbl:BLParametersComparison}.
\begin{table}
\caption{Boundary layer parameters as computed via the current
  temporal slow growth approach, the spatial slow growth method
  of~\citet{spalart1988direct}, and a spatially evolving
  simulation~\citep{schlatter2010assessment}.}
\centering
\begin{tabular}{lccccccc}
\hline
  Method    & Re$_\theta$ & $H = \delta^*/\theta$ & $c_f$ \\
\hline
  Temporal slow growth &  703 & 1.49  & \num{4.70e-3} \\
  Spatial slow growth & 670 & 1.49 & \num{4.86e-3} \\
  Spatially evolving & 677 & 1.47 & \num{4.78e-3} \\
\hline
\end{tabular}
\label{tbl:BLParametersComparison}
\end{table}

\begin{figure}[htp]
\begin{center}
\includegraphics[width=0.7\linewidth]{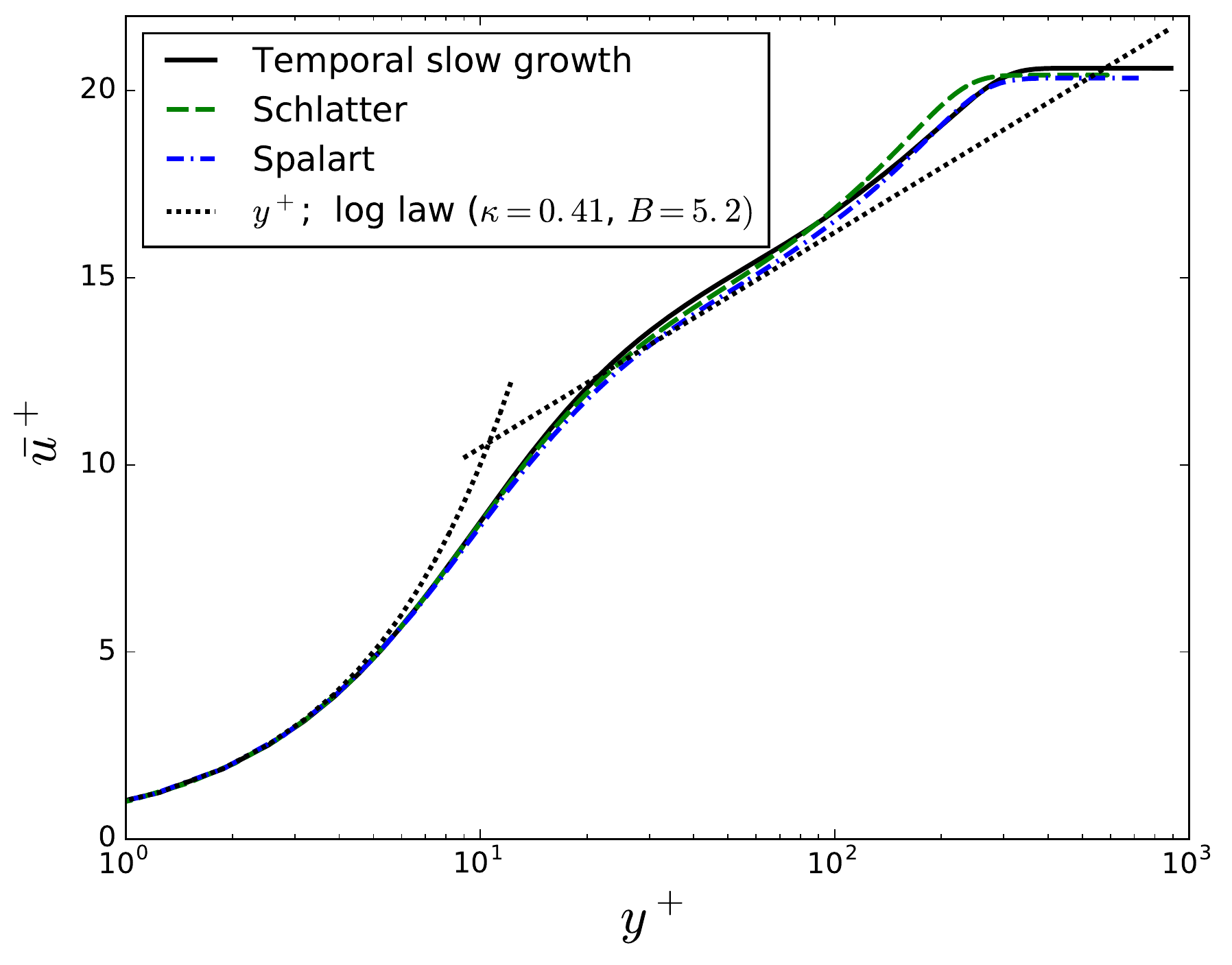}\\
\includegraphics[width=0.7\linewidth]{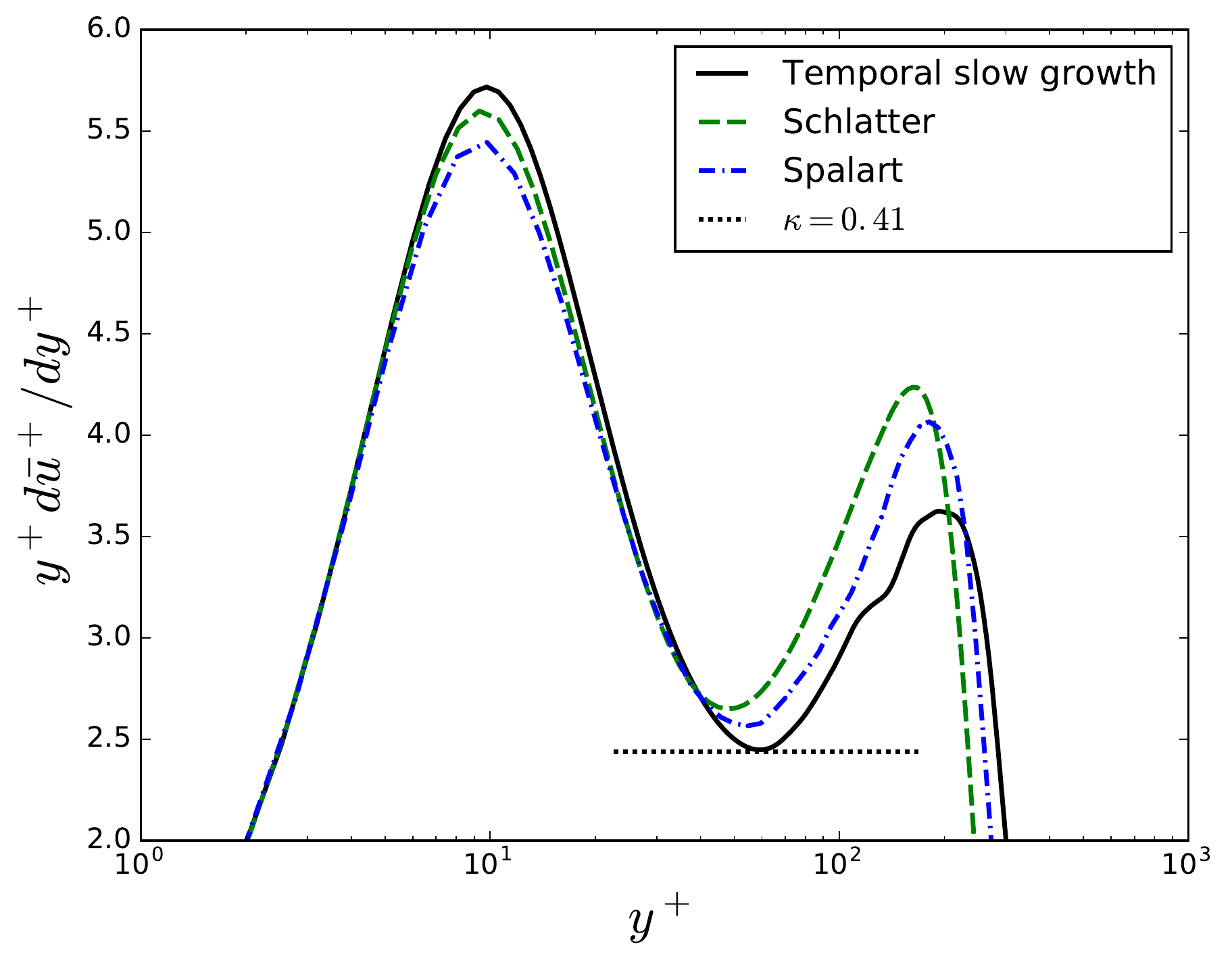}
\end{center}
\caption{Mean streamwise velocity and its derivative, normalized by
  the viscous scales. \label{fig:L_Ueff}}
\end{figure}

Figure~\ref{fig:L_Ueff} shows the mean streamwise velocity $U^+$ and
the quantity $\beta=y\partial U^+/\partial y$, which, in a log layer,
will be constant with value $1/\kappa$, where $\kappa$ is the Karman
constant. Curves for the law of the wall in the viscous sublayer
($U^+=y^+$) and in the logarithmic layer ($U^+=\log(y^+)/0.41+5.2$) are
also shown.
The mean velocity in the temporal DNS is qualitatively similar to that
of both spatial simulations and follows closely the linear and
logarithmic profiles.  However, examining the quantity $\beta$ makes
clear that there is not really a region over which the velocity varies
logarithmically in any of the three simulations, because the Reynolds
numbers are much too low. In channel flow, an order of magnitude larger
Reynolds number was required to observe a significant logarithmic
region \citep{LeeMoser2015}.  In the temporal case, the minimum of
$\beta$ occurs with a value corresponding to $\kappa=0.41$, which is
slightly larger than that observed for the spatial
simulations. However, the simulations of \citet{LeeMoser2015} indicate
that the value of $\kappa$ in an actual log layer at higher Reynolds
number is likely to be significantly lower than this, since in the
channel flow simulation, the minimum in $\beta$ is about 15\% lower
than the value in the log region.

%

\begin{figure}[p]
\begin{center}
  \subfigure[Total shear stress.]{\includegraphics[width=0.5\linewidth]{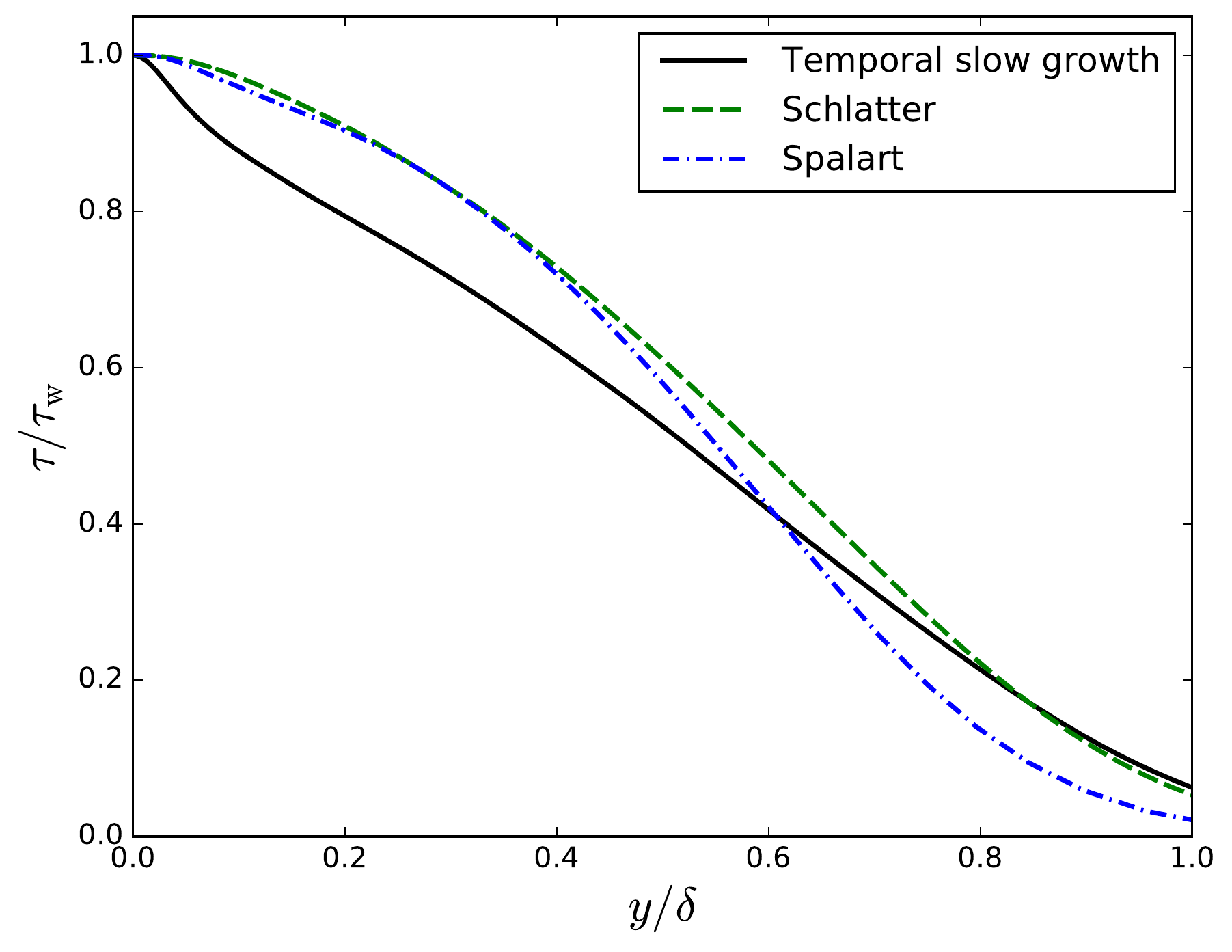}}
  \subfigure[Viscous shear stress.]{\includegraphics[width=0.5\linewidth]{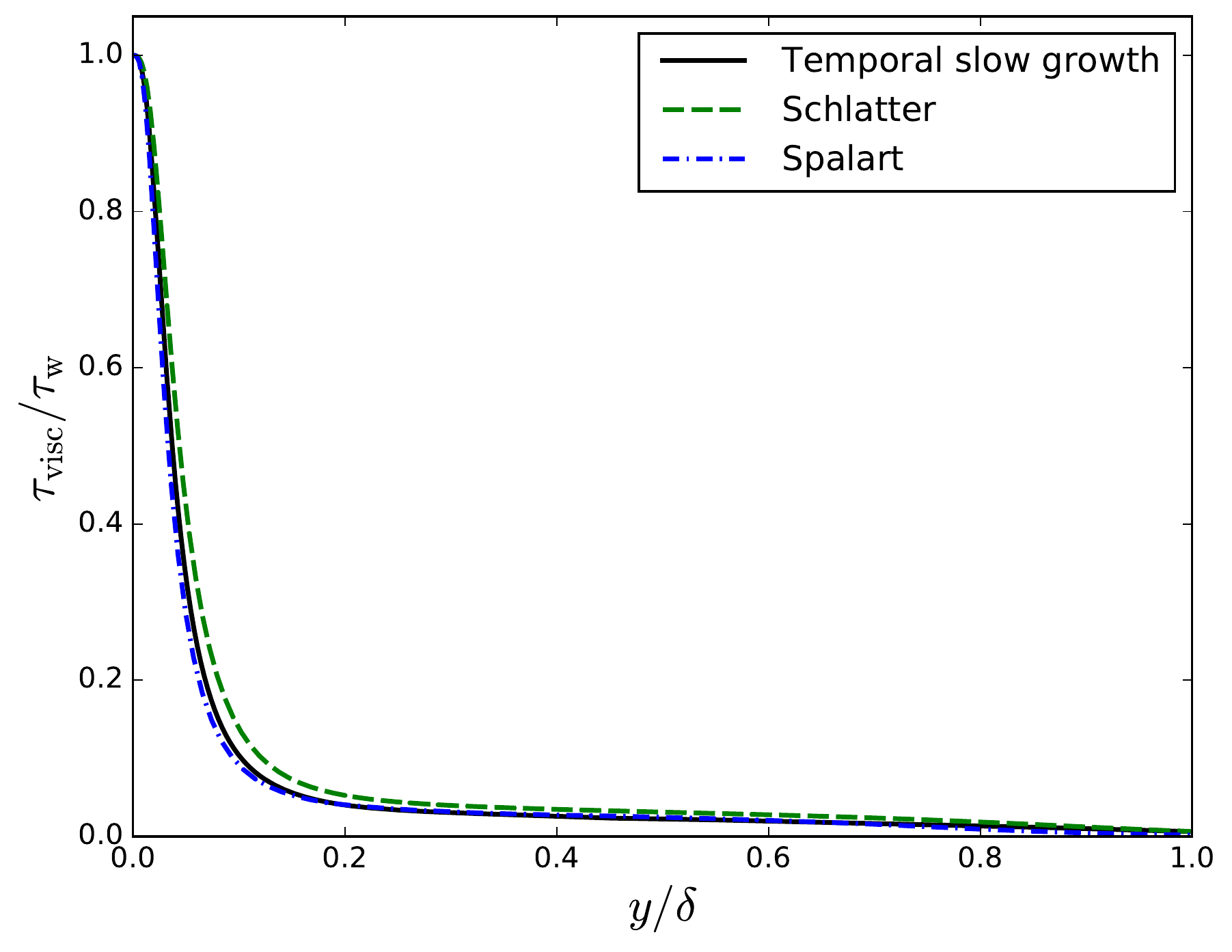}}
  \subfigure[Turbulent shear stress.]{\includegraphics[width=0.5\linewidth]{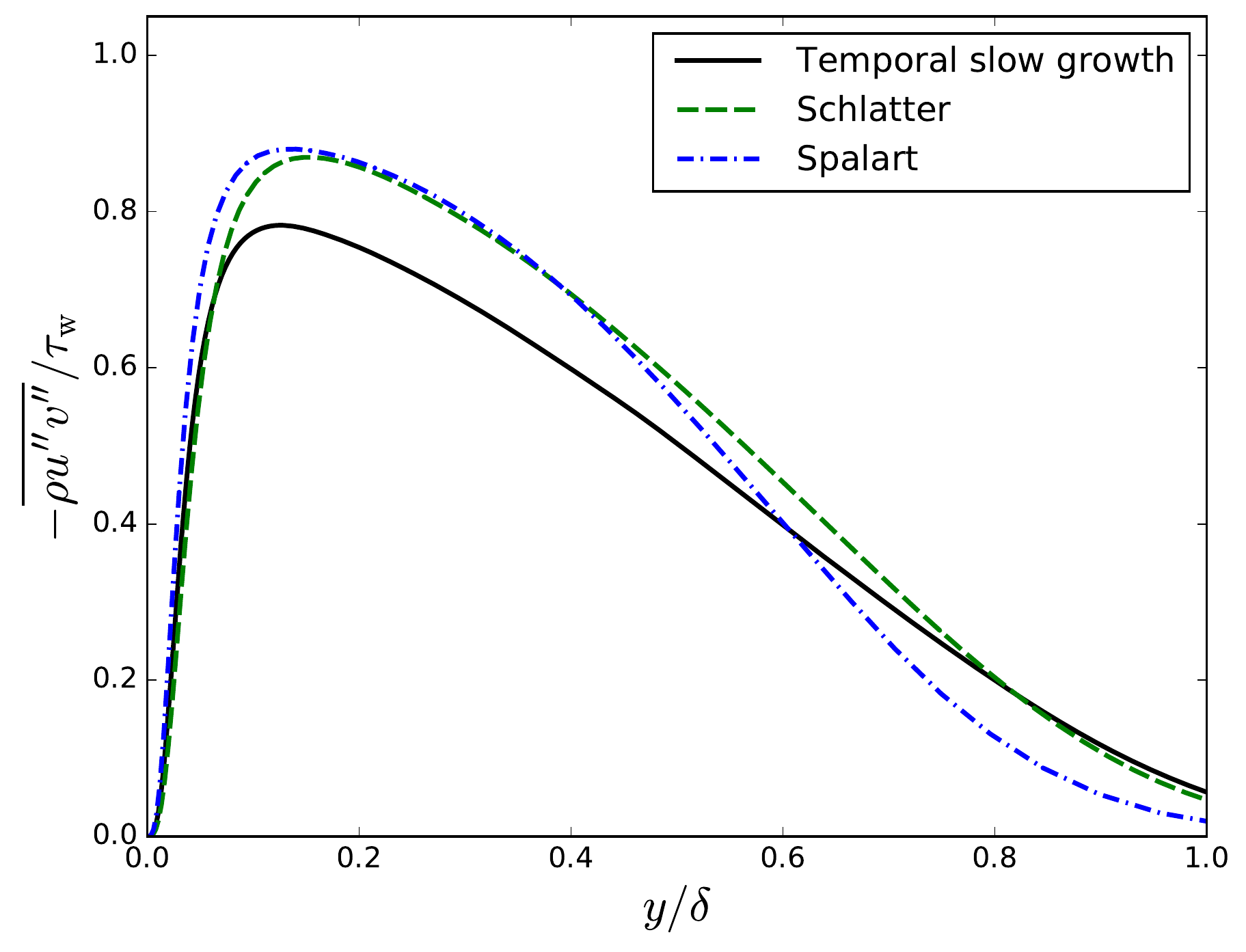}}
\end{center}
\caption{Shear stresses, normalized by the shear stress at the wall. \label{fig:L_shear}}
\end{figure}

Figure~\ref{fig:L_shear} shows the mean shear stress normalized by the wall
shear stress.
The shape of the total shear stress in the temporally homogenized
boundary layer differs qualitatively from both the
spatially homogenized and spatially evolving cases.  In particular,
as expected, the derivative of the total shear stress is zero
at the wall in all cases, but the stress drops more quickly in the buffer
layer in the temporally homogenized boundary layer.  The mean viscous stress is
essentially the same for the three different models, as expected given
the mean velocity.  Thus, the difference in the total stress is due to
the Reynolds shear stress, with the peak value in the
temporally homogenized simulation approximately 10\% lower than
in either of the spatial cases.

The observed differences in the behavior of the total shear stress
can be explained by examining the relationship between the
total stress and the mean velocity implied by the boundary layer
approximation of the mean momentum equation.  In particular, in a
spatially evolving, zero-pressure-gradient, constant-density boundary
layer, the boundary layer equations imply that the total shear stress
is given by
\begin{equation*}
\frac{\tau}{\tau_\wall}
= 
1 + \frac{\nu}{u_{\tau}^2} \dd{u_{\tau}}{x} \int_0^{y^+} (u^+)^2 \diff y^+.
\end{equation*}
Alternatively, the temporal slow growth formulation leads to
\begin{equation*}
\frac{\tau}{\tau_\wall} = 1 - \grt(\Delta)^+  \left[ u^+ y^+ - \int_{0}^{y^+} u^+ \diff y^+ \right],
\end{equation*}
where $\grt(\Delta)^+ = \nu \grt(\Delta) / u_{\tau}^2$.  These forms
behave differently near the wall, leading to the discrepancies in
total shear and Reynolds shear stress shown in
Figure~\ref{fig:L_shear}.  For example, in the viscous sublayer where
$u^+ = y^+$, the spatially evolving result is $\tau /\tau_\wall = 1 -
C_s (y^+)^3$, while the temporal slow growth boundary layer gives $\tau /
\tau_\wall = 1 - C_t (y^+)^2$, where $C_s$ and $C_t$ are
problem-dependent, positive constants.  The temporal slow growth
behavior in the viscous sublayer is consistent with a temporally
evolving boundary layer---see Appendix~\ref{app:total_stress} for
more details---which leads to the observed discrepancies between the
total stress in the current simulations and the spatially homogenized
or evolving cases.

\begin{figure}[p]
\begin{center}
\subfigure[Streamwise velocity RMS.]{\includegraphics[width=0.5\linewidth]{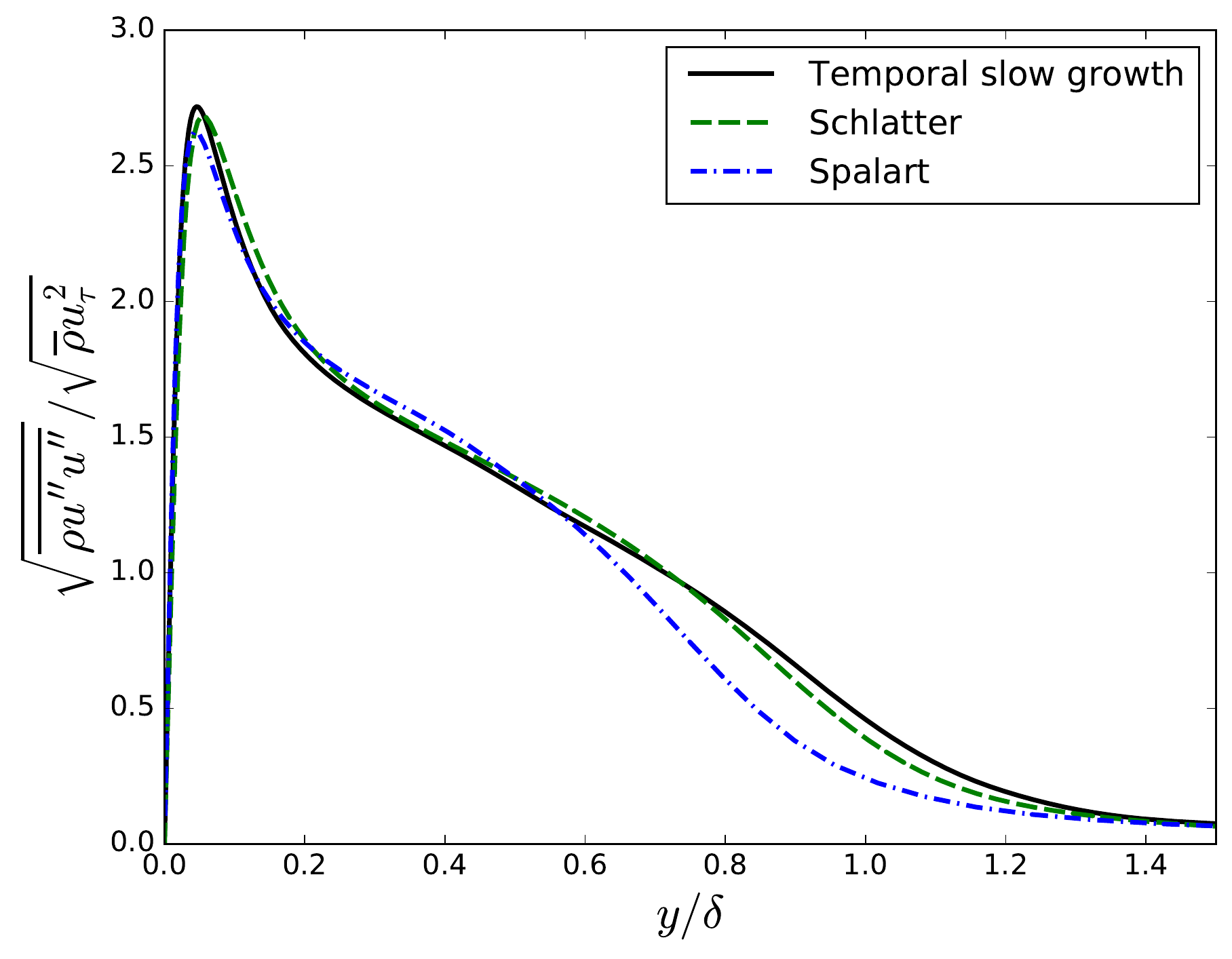}}
\subfigure[Wall normal velocity RMS.]{\includegraphics[width=0.5\linewidth]{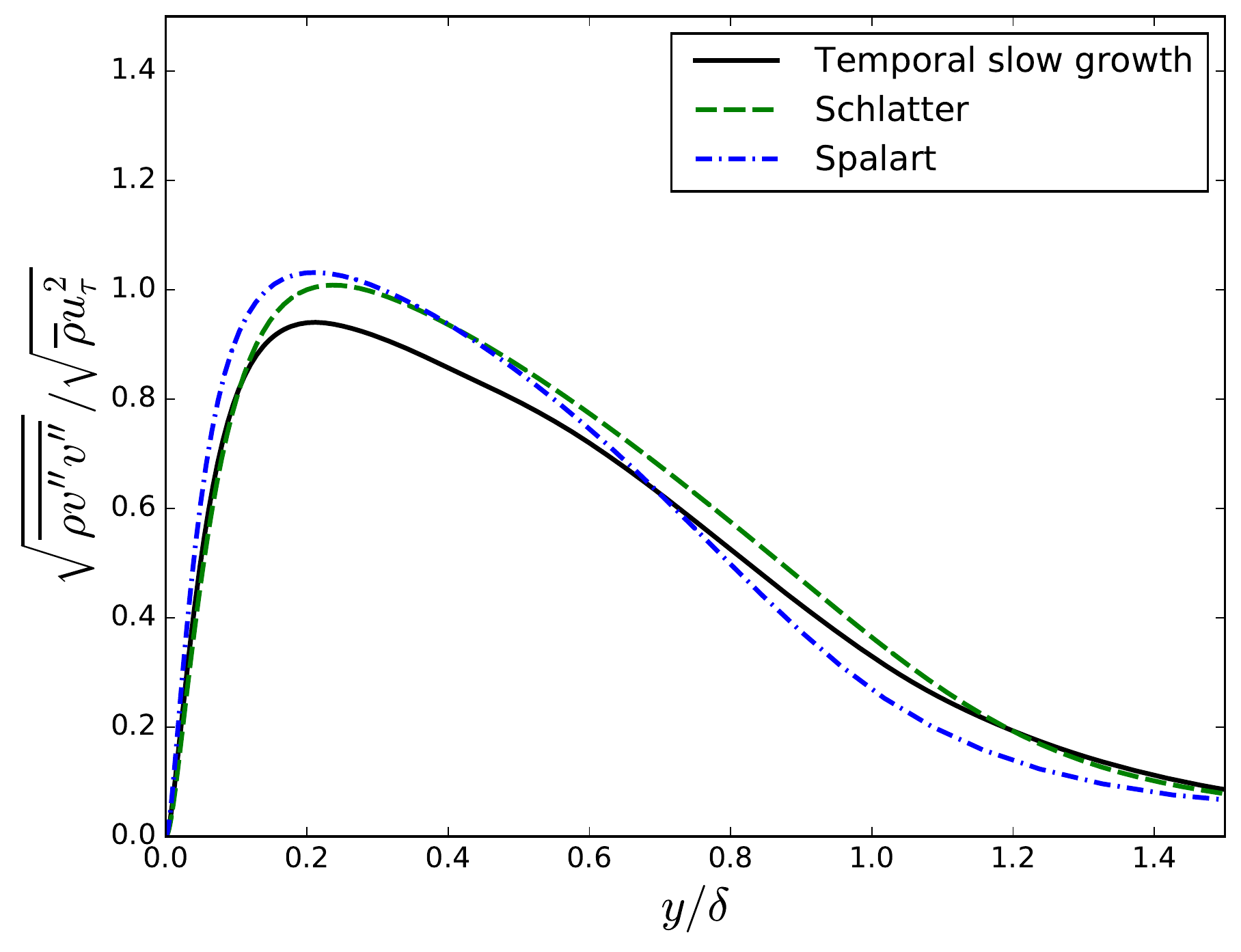}}
\subfigure[Spanwise velocity RMS.]{\includegraphics[width=0.5\linewidth]{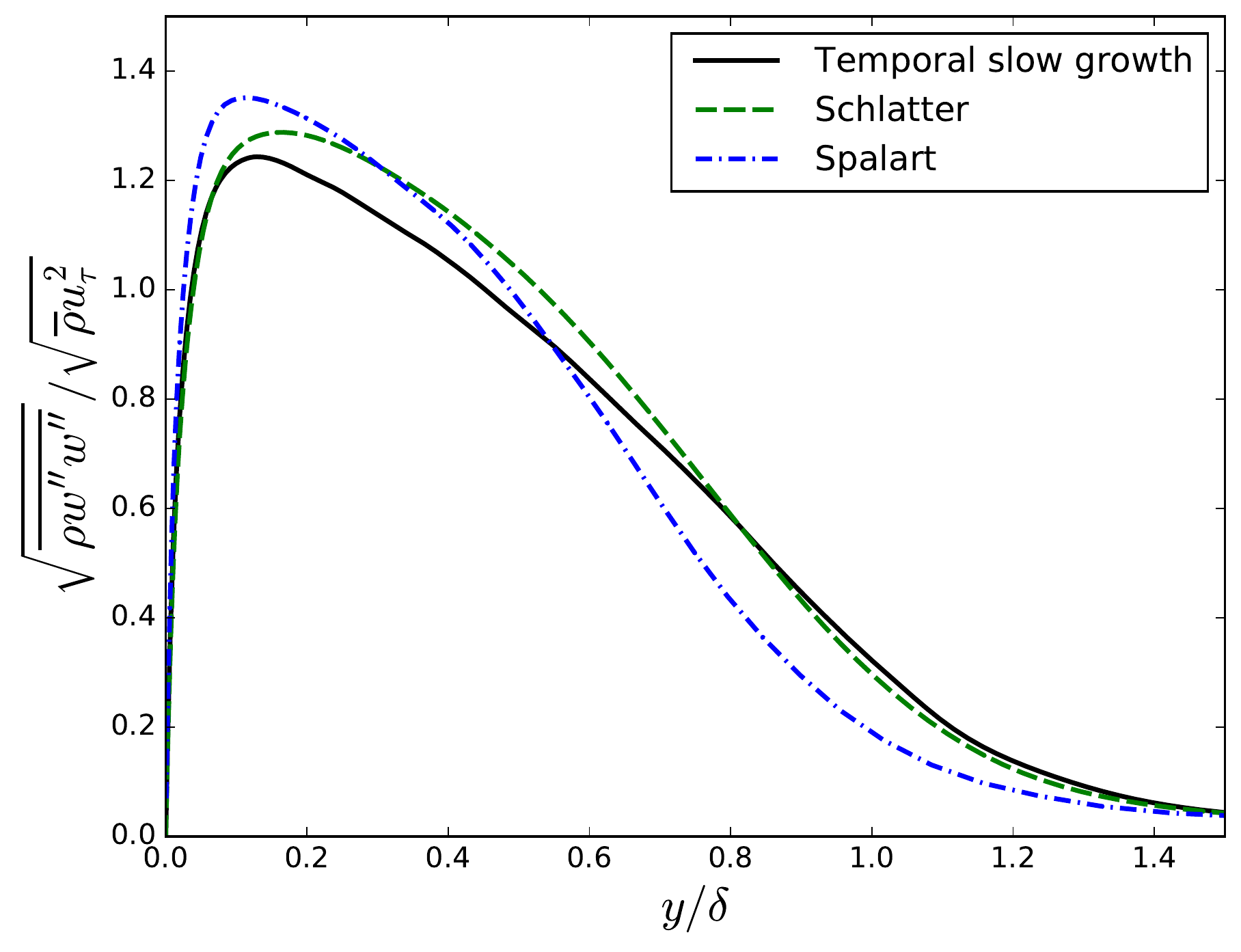}}
\end{center}
\caption{RMS velocity components. \label{fig:L_velRMS}}
\end{figure}

Figure~\ref{fig:L_velRMS} shows the RMS of the velocity fluctuations
normalized by $u_{\tau}$.
As for the shear stresses, there is a reasonable agreement between the
three simulations for the RMS velocities.  The streamwise component shows
particularly good agreement, with both the location and magnitude of
the peak in close agreement between all three simulations.  For the
wall-normal and spanwise components, the temporal slow growth results
tend to be below the spatial simulations, with the discrepancy near
the peak being roughly 10\%.  Near the boundary layer edge, say for
$y/\delta > 0.8$, the temporal slow growth RMS velocities all agree
better with the spatially evolving case than do the spatially
homogenized profiles, although it is unclear why.

\begin{figure}[htp]
\begin{center}
\subfigure[Near wall region (non-dimensionalized by $u_{\tau}$ and $\nu/u_{\tau}$).]{\includegraphics[width=0.6\linewidth]{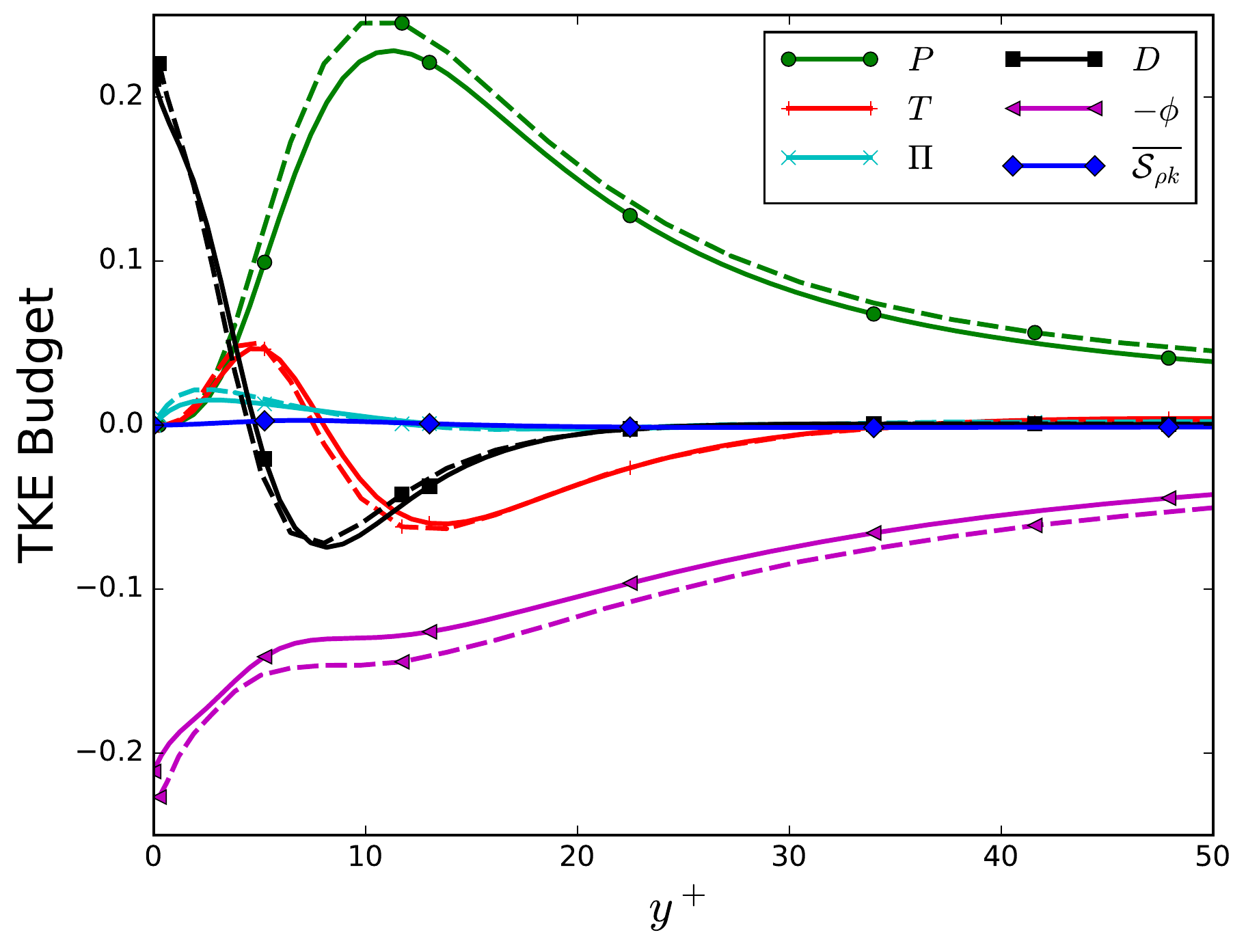}}
\subfigure[Outer region (non-dimensionalized by $u_{\tau}$ and $\delta$).]{\includegraphics[width=0.6\linewidth]{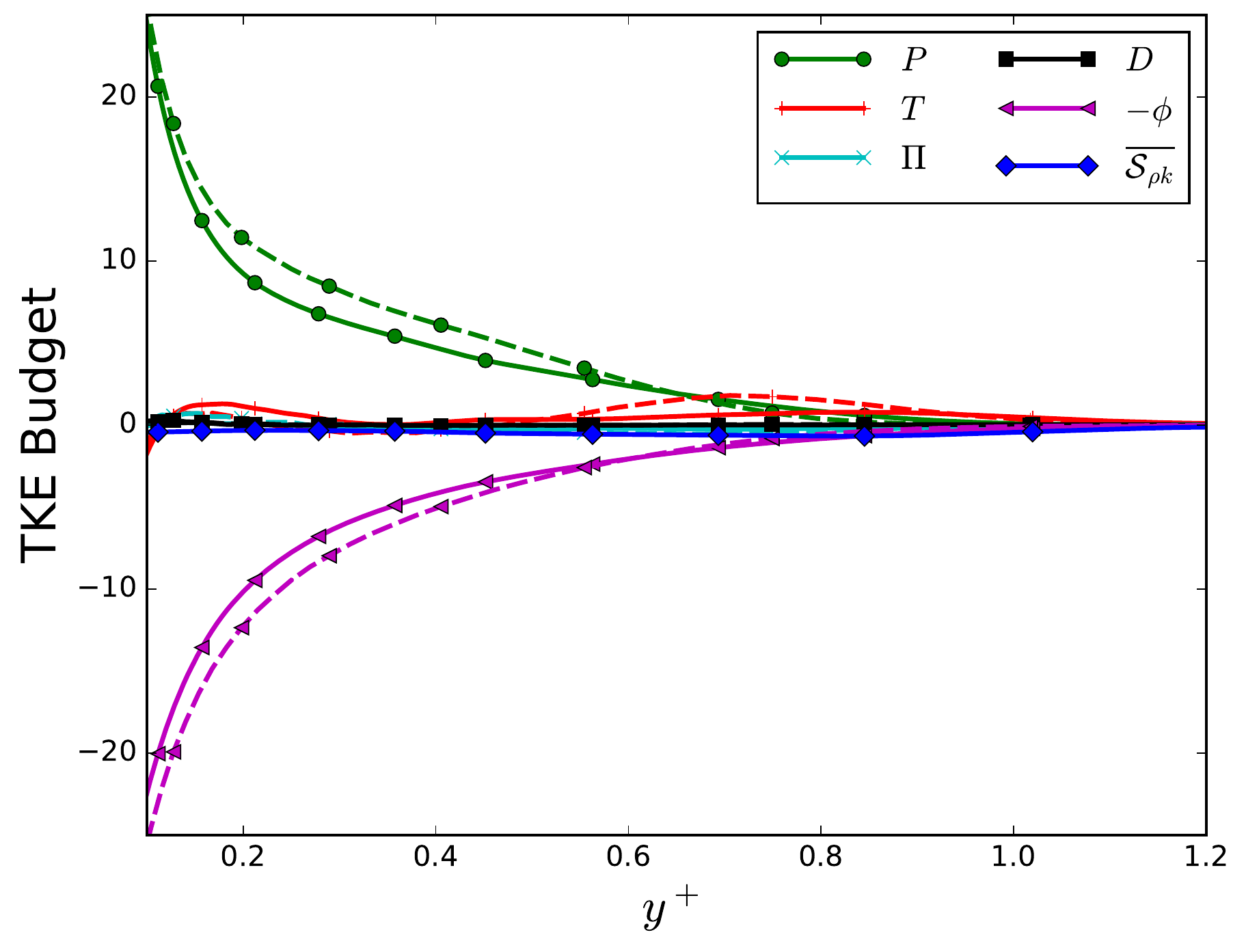}}
\end{center}
\caption{Turbulent kinetic energy budget. Solid lines show results
  from the temporal slow growth approach.  Dashed lines show results
  from spatial slow growth approach
  of~\citet{spalart1988direct}. \label{fig:L_tke_budget}}
\end{figure}

The turbulent kinetic energy budget is shown in
Figure~\ref{fig:L_tke_budget}.
Specifically, using homogeneity in the streamwise ($x_1$) and spanwise
($x_3$) directions, the TKE equation can be written
\begin{equation*}
\pp{\bar{\rho} k}{t}
=
C + P + T + \Pi + D - \phi + V + \overline{\Ssd_{\rho k}},
\end{equation*}
where, in index notation,
\begin{gather*}
C = -\, \tilde{u}_2 \pp{\rho k}{x_2}, \quad
P = - \overline{\rho u''_2 u''_i} \pp{\tilde{u}_i}{x_2}, \quad
T = -\,\frac{1}{2} \pp{}{x_2} \left( \overline{\rho u''_i u''_i u''_2} \right), \\
\Pi = -\, \pp{}{x_2} \left( \overline{u''_2 p'} \right) + \overline{p' \pp{u''_i}{x_2} }, \quad
D = \pp{}{x_2} \left( \overline{u''_i \tau'_{i2}} \right), \quad
\phi = \overline{\tau'_{ij} \pp{u''_i}{x_j} }, \\
V = -\,\overline{u_2''} \pp{\bar{p}}{x_2} + \overline{u_i''} \pp{\bar{\tau}_{ij}}{x_j} - \bar{\rho} k \pp{\tilde{u}_2}{x_2}, \quad
\overline{\Ssd_{\rho k}} = x_2 \grt(\Delta) \pp{\bar{\rho} k}{x_2}.
\end{gather*}
Because the density is essentially constant in this case, for the
temporal formulation, $\tilde{u}_2 \approx 0$, which implies that the
mean convection term $\tilde{u}_2 \partial (\bar{\rho} k) / \partial
x_2$ is negligible.  Further, the compressibility terms in $V$ are
also negligible.  Thus, only $P$, $T$, $\Pi$, $D$, $\phi$, and
$\overline{\Ssd_{\rho k}}$ are shown.

Near the wall, the features of the dominant terms in the
TKE balance from the temporal slow growth simulation are similar to
those from the spatial slow growth case.  Both production and
dissipation are somewhat smaller in the temporal case, which is
consistent with the reduced Reynolds shear stress observed in
Figure~\ref{fig:L_shear}.  The viscous and turbulent transport terms
however match almost perfectly.  Away from the wall, production and
dissipation remain smaller in the temporal simulation, and the outer
peak in the turbulent transport is significantly reduced.

To summarize, the temporal slow growth model flow mimics many of the
important features of the statistics of a zero-pressure-gradient,
spatially evolving boundary layer.  The mean velocity, streamwise RMS
velocity, and dominant near-wall terms in the $k$ budget are
particularly well-represented.  However, as expected, the differences
between temporal and spatial evolution of the boundary layer and the
approximations inherent to the slow-growth formulation lead to
some obvious discrepancies.
For instance, the Reynolds shear stress, wall-normal RMS velocity, and
spanwise RMS velocity are all lower in the temporal simulation than in
the spatially homogenized or spatially evolving cases.  Such
differences are relevant if the goal is to investigate the
characteristics of a truly spatially evolving boundary layer; however,
they do not diminish the utility of temporally homogenized boundary
layers for studying wall-bounded turbulence more generally or for
RANS model evaluation, as discussed in Section~\ref{sec:introduction}.


\subsection{Case (\cev): $M_{\infty}=1.2$, Cold Wall Boundary Layer with Transpiration}
\label{sec:results_highM}

\begin{figure}[htp]
\begin{center}
\includegraphics[width=0.6\linewidth]{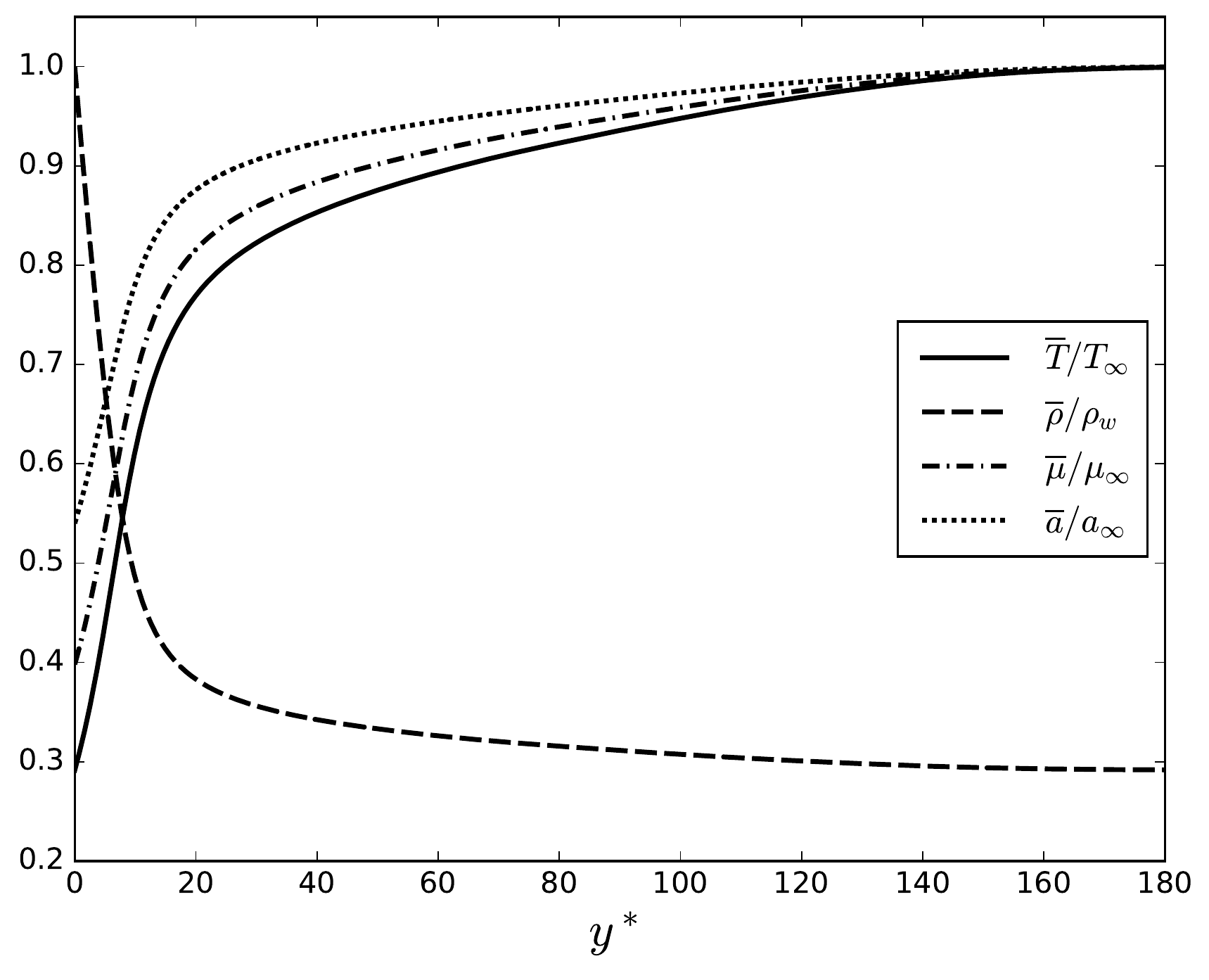}
\includegraphics[width=0.6\linewidth]{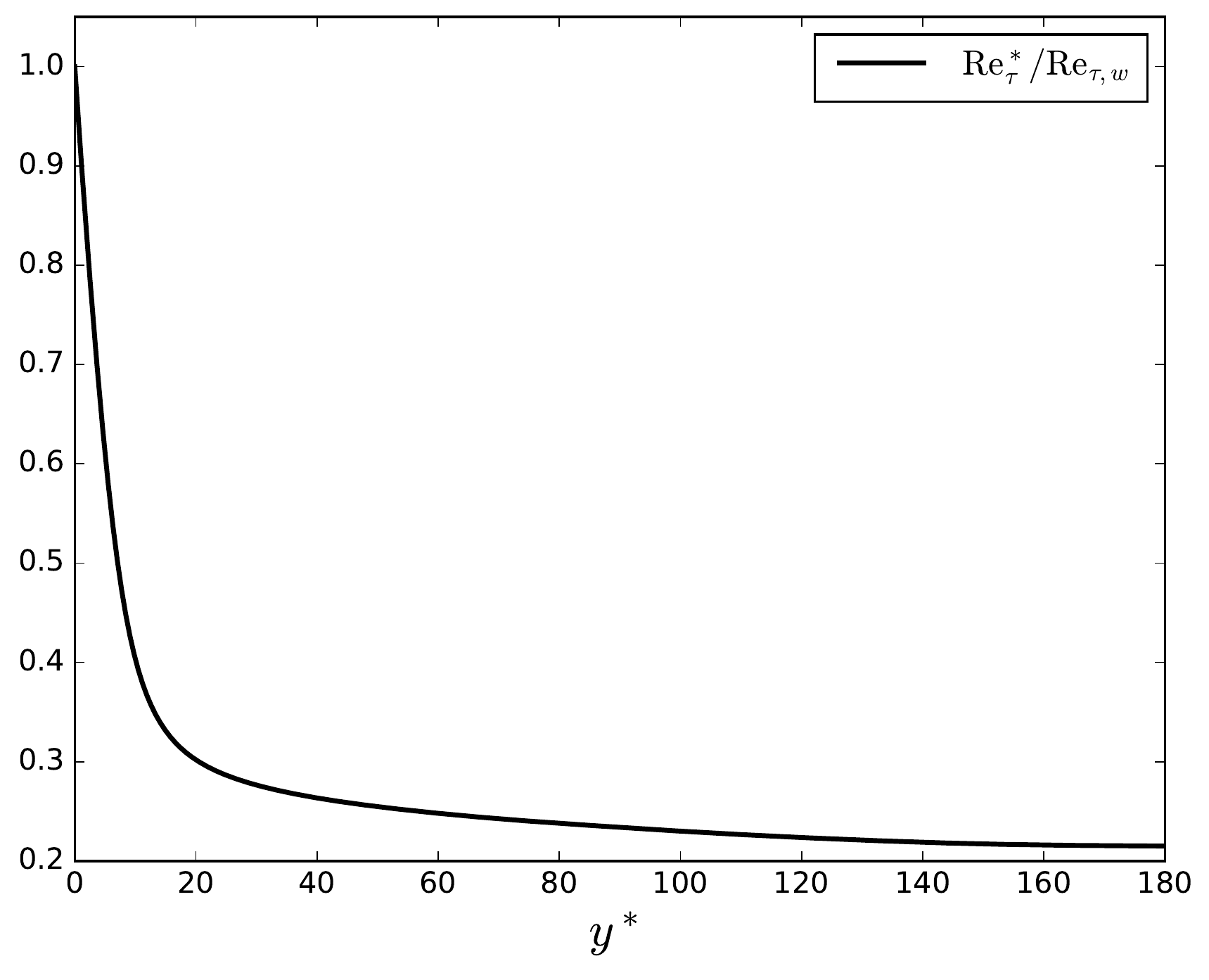}
\end{center}
\caption{Variation of thermodynamic quantities and viscosity as a
  function of wall distance normalized by the semi-local length scale
  (top) and ratio of Reynolds number based on semi-local friction
  velocity, boundary layer thickness, and local kinematic viscosity
  relative to Reynolds number based on friction velocity, boundary
  layer thickness, and wall kinematic viscosity. \label{fig:C_thermo}}
\end{figure}

Results from the Case (\cev) simulation are presented and compared
with those from the Case (\lowM) in this section.  Many of the statistics
are normalized using the semi-local scaling introduced by
\citet{morinishi2004direct}, where local mean viscosity and density
are used in the friction velocity and viscous length scale rather than
wall values.  Hence, the semi-local friction velocity is
$u_{\tau^*}=\sqrt{\tau_\wall / \overline{\rho}}$, and the semi-local
viscous length scale is
$\delta_{\nu^*}=\overline{\mu}/(\overline{\rho} u_{\tau^*})$.  The
wall distance normalized by the semi-local viscous scale is denoted
$y^*=y/\delta_{\nu^*}$.  The use of this scaling has almost no effect
on the Case (\lowM) profiles. In Case (\cev), the use of this scaling
is justified by the strong variation in density and viscosity near the
wall due to the cold wall.  As is evident in
Figure~\ref{fig:C_thermo}, most of the variation in mean thermodynamic
and transport quantities occurs in the viscous sublayer and buffer layer
where $y^* \leq 20$.
This strong variation in mean properties leads to a large variation in
local Reynolds number across the boundary layer, with the near-wall
region having the highest Re based on local properties.

\begin{figure}[htp]
\begin{center}
\includegraphics[width=0.6\linewidth]{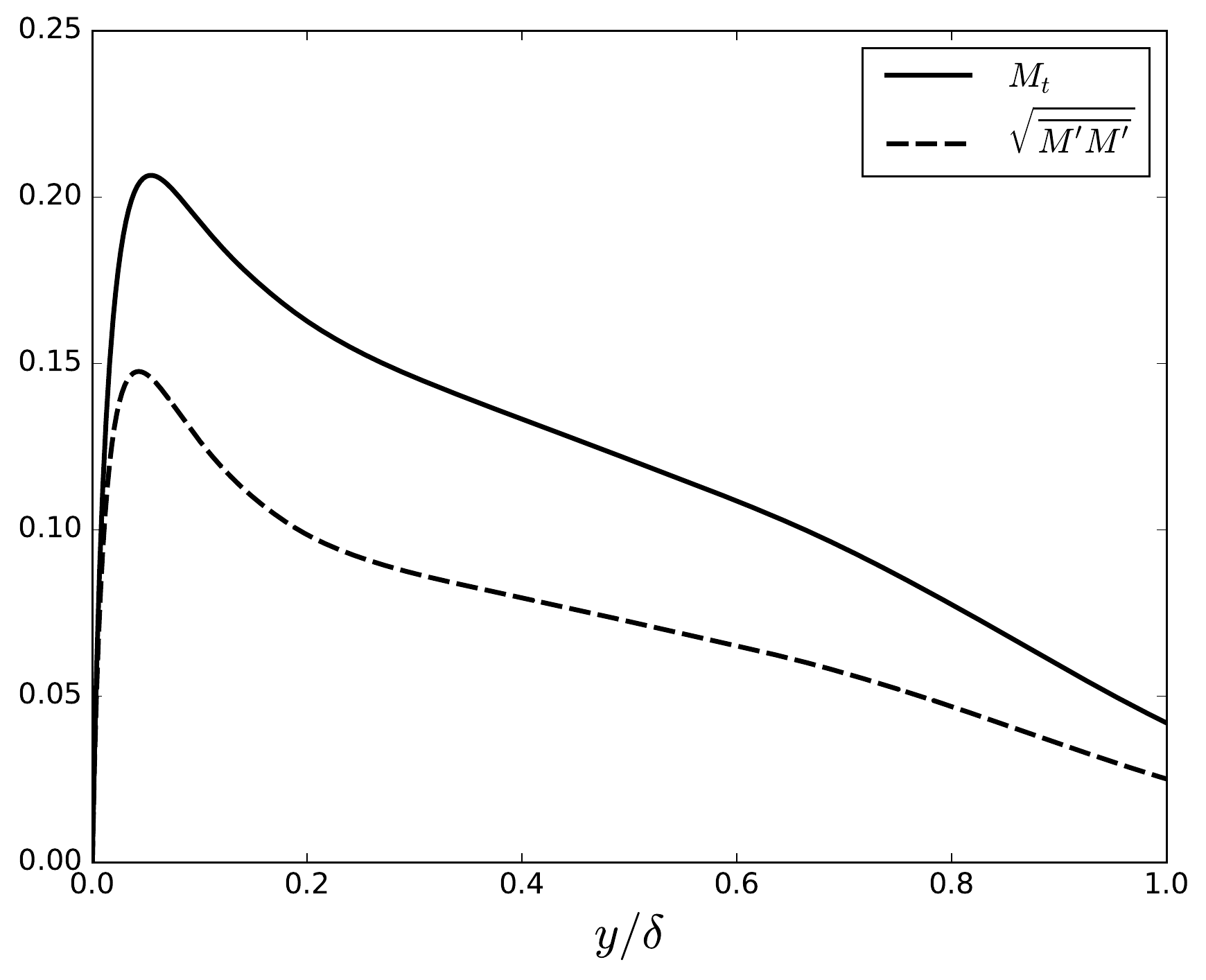}
\end{center}
\caption{Turbulent Mach number and Mach number RMS. \label{fig:C_Mt}}
\end{figure}

While the thermodynamic and transport properties vary dramatically,
the turbulent Mach number $M_t = \sqrt{\widetilde{u''_i
    u''_i}}/\overline{a}$, shown in Figure~\ref{fig:C_Mt}, is low,
with a maximum of approximately 0.2, as expected in a mildly
supersonic boundary layer.
Therefore, according to Morkovin's hypothesis~\citep{Morkovin1962,
  Smits2006}, it is expected that the effects of compressibility on
turbulence are very weak for this case, although the property
variations will cause the results to differ substantially from a
low Mach boundary layer.

\begin{figure}[htp]
\begin{center}
\includegraphics[width=0.9\linewidth]{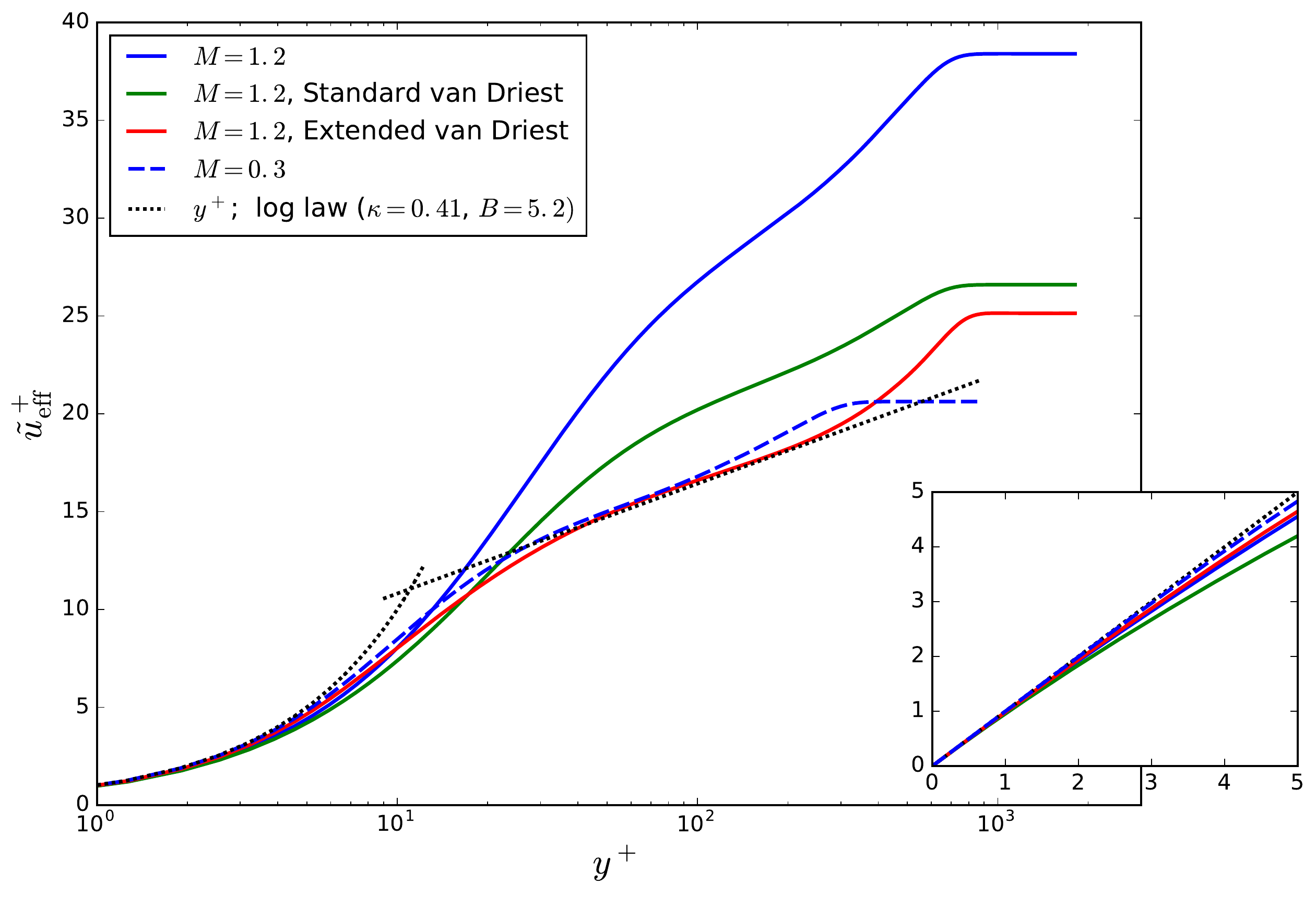}
\end{center}
\caption{Mean streamwise velocity.  Five profiles are shown: The raw
  Case (\cev) profile (solid blue), the van Driest transformed Case
  (\cev) profile (solid green), the extended van Driest transformed
  (see Appendix~\ref{app:evd}) Case (\cev) profile (solid red), the
  Case (\lowM) profile (dashed blue), and the law of the wall (dotted
  black).  The inset shows the viscous sublayer only.\label{fig:C_Ueff}}
\end{figure}

Figure~\ref{fig:C_Ueff} shows the mean velocity for Case (\cev).
Three different transformations of the Case (\cev) streamwise mean
velocity are shown.  The first, shown in blue, is simply the mean
velocity normalized by the friction velocity.  Of course, this
normalization does not account for variable property effects and, as
expected, the result does not collapse on the Case (\lowM) profile
(blue dashed line) or the incompressible law of the wall (black dotted
line).  It is common practice to consider the~\citet{vanDriestTrans}
transformed mean velocity when comparing compressible boundary layers
to their incompressible counterparts, and this transformation is often
successful in collapsing the
profiles~\citep{White_1991_Viscous_Flow}.  The van Driest
transformed velocity is shown in Figure~\ref{fig:C_Ueff} in green.
While this profile is closer to the incompressible velocity profile
than the untransformed velocity, there are still substantial
discrepancies.  First, the transformed velocity is below the
$u^+ = y^+$ curve for $y^+$ less than 3, as is clear in the inset of
Figure~\ref{fig:C_Ueff}.
Second, there is a large offset in the log layer, which would
lead to a log layer offset constant greater than 10.  These
discrepancies can be explained by the fact that the transformation
does not account for the effects of wall transpiration or a highly
cooled wall.  The effects of the cold wall have been previously
examined by~\citet{huang1994van}, who proposed a modified
transformation that accounts for the temperature variation in
the viscous sublayer.  In Appendix~\ref{app:evd}, we develop a further
extension of the transformation of~\citet{huang1994van} that accounts
for both the cold wall and wall transpiration.  The result of applying
this transformation is shown in red in Figure~\ref{fig:C_Ueff}.
The transformation is quite successful in collapsing the Case
(\cev) velocity profile with incompressible theory, indicating that mean
property variation accounts for the differences between the
compressible and incompressible mean velocity profiles for this case.

%

\begin{figure}[htp]
\begin{center}
\subfigure[Inner region.]{\includegraphics[width=0.6\linewidth]{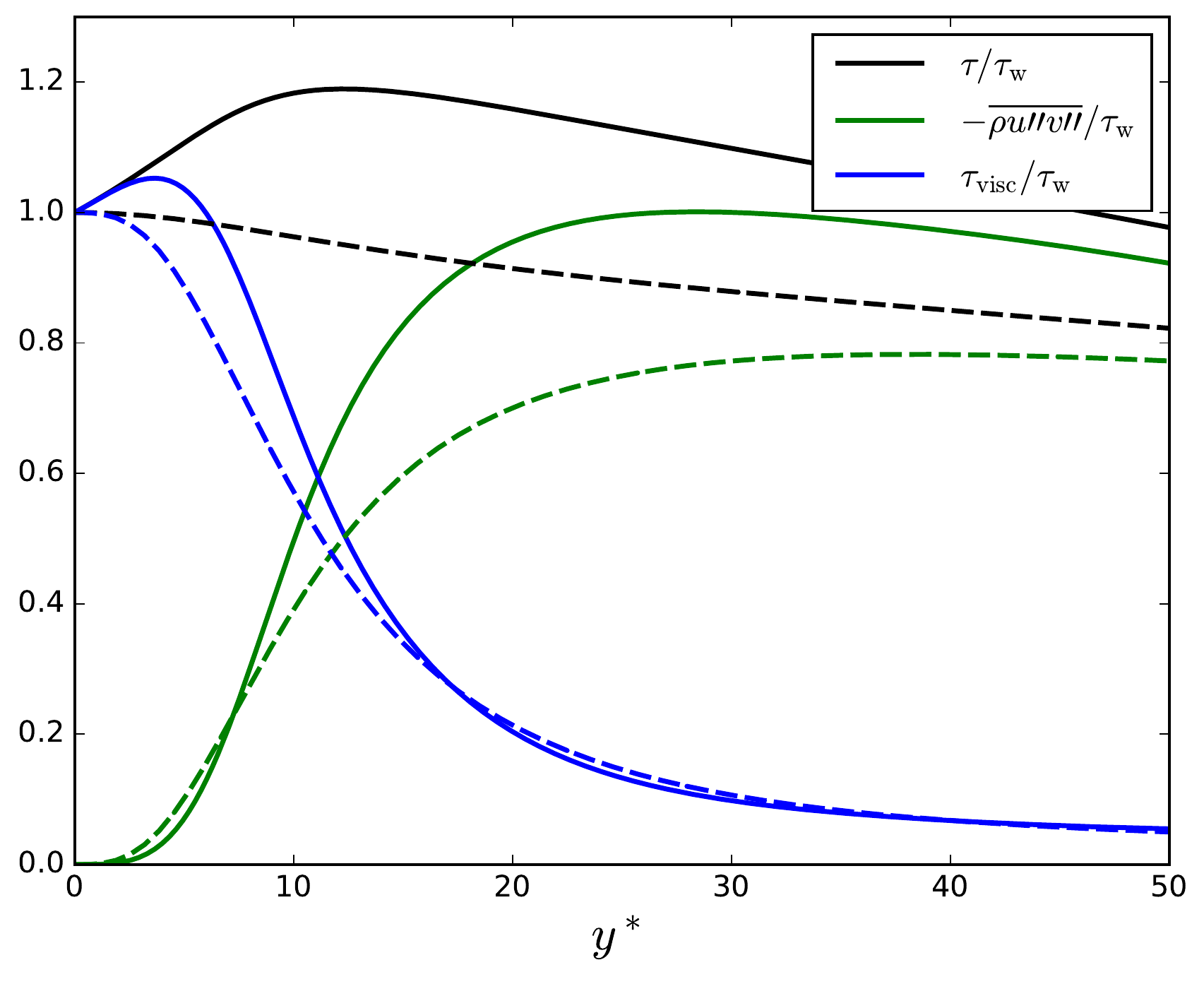}}
\subfigure[Entire boundary layer.]{\includegraphics[width=0.6\linewidth]{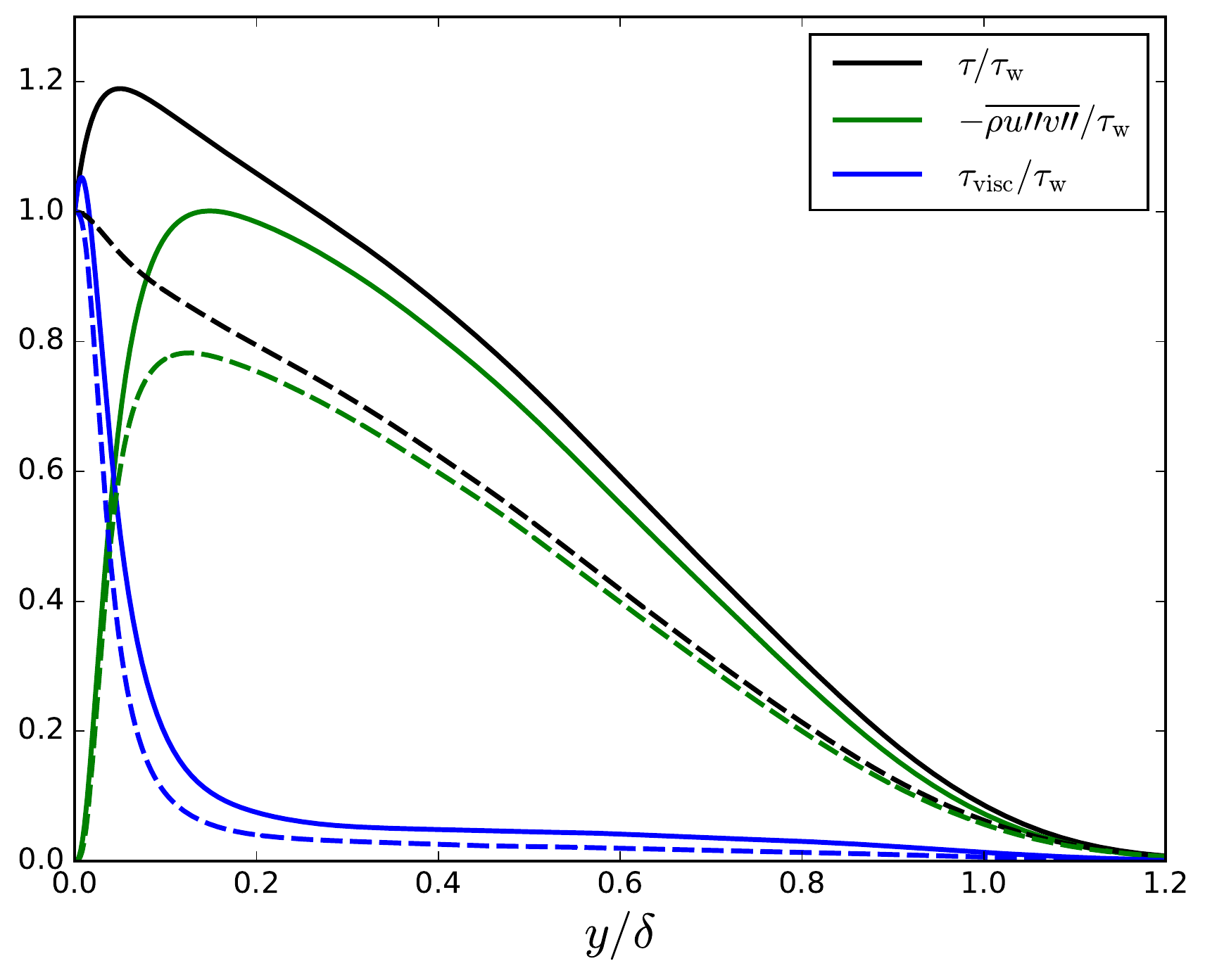}}
\end{center}
\caption{Shear stress components, normalized by the wall stress.  The
  solid lines represent Case (\cev) while the dashed lines are from
  Case (\lowM). \label{fig:C_shear}}
\end{figure}

Figure~\ref{fig:C_shear} shows the shear stresses.
Unlike Case (\lowM), the total shear stress has a positive derivative
at the wall and peaks near $y^{*} \approx 12$ with a value approximately $20\%$
larger than at the wall.  These features are a consequence of
the wall transpiration.  With wall transpiration, the term $\bar{\rho} \tilde{v}
\partial \tilde{u} / \partial y$ in the mean momentum equation, which
is zero at the wall and negligible near the wall in the non-blowing
case, is non-zero even at the wall.  This term leads to a larger total shear
over the entire boundary layer, which, outside of the viscous
sublayer, leads to a larger Reynolds shear stress.  In particular, at
its peak, the Reynolds shear stress is also approximately $20\%$
larger in Case (\cev) than in Case (\lowM).

\begin{figure}[htp]
\begin{center}
\subfigure[Inner region.]{\includegraphics[width=0.6\linewidth]{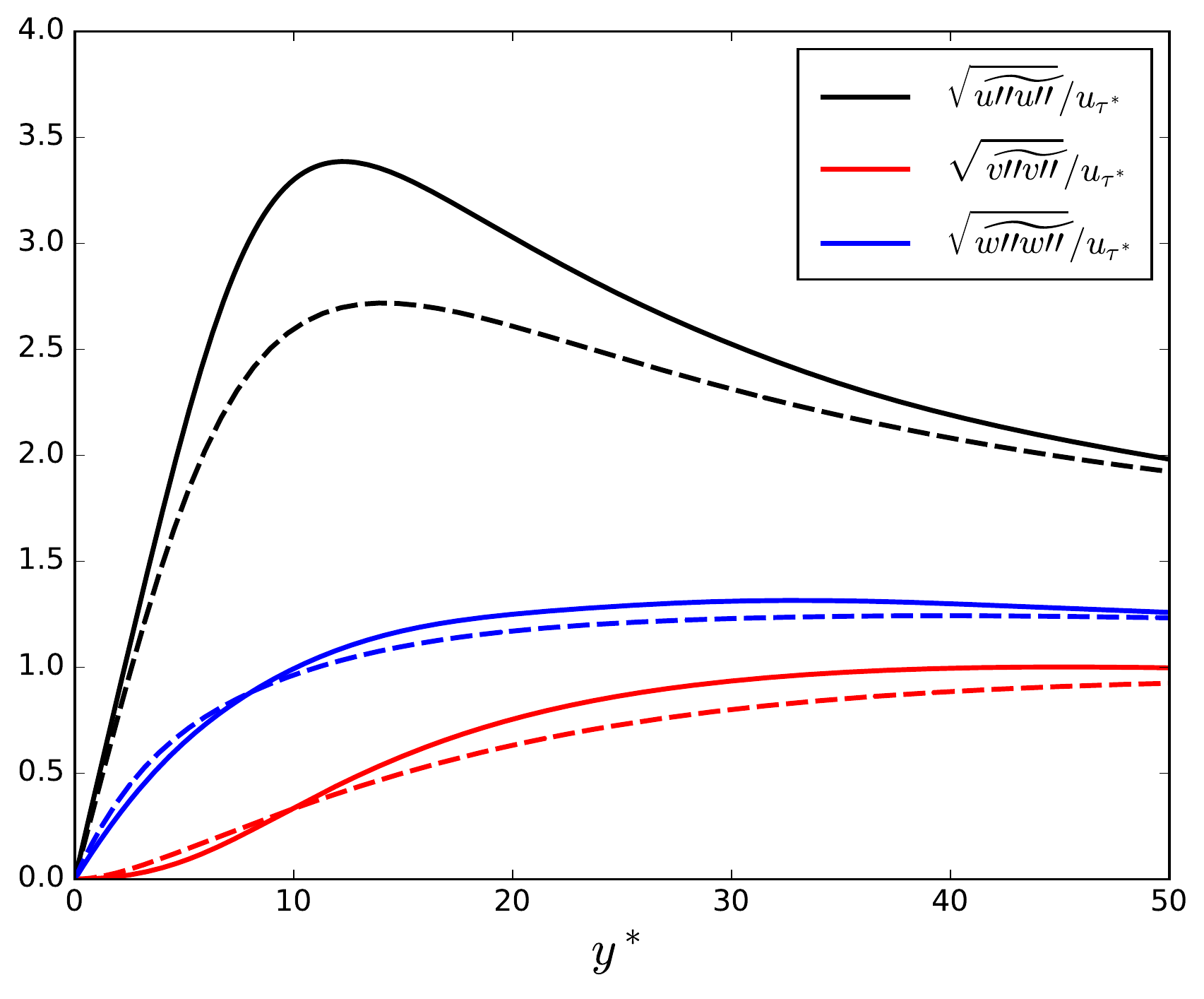}}
\subfigure[Entire boundary layer.]{\includegraphics[width=0.6\linewidth]{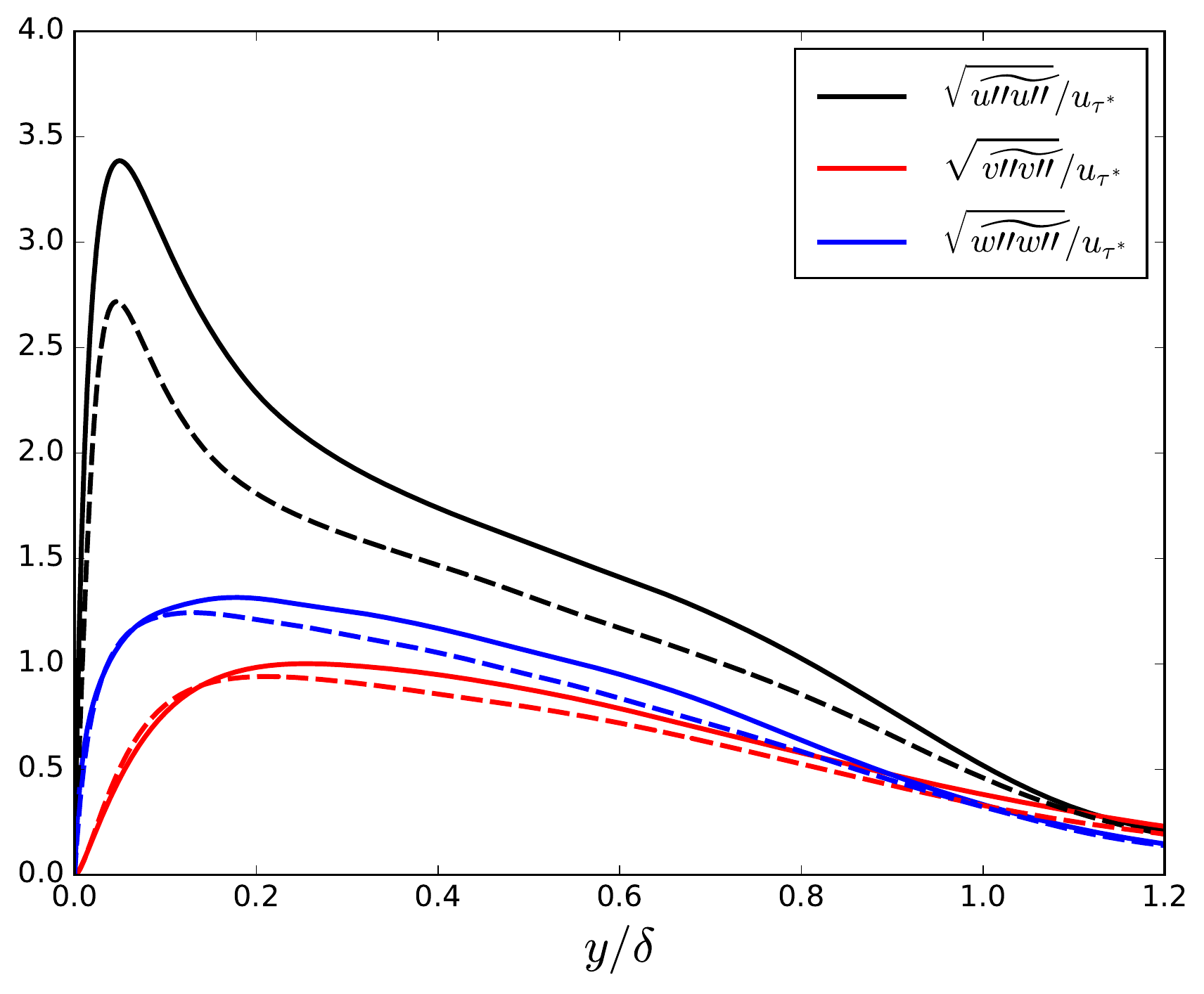}}
\end{center}
\caption{RMS velocities, normalized by the semi-local friction
  velocity.  The solid lines represent Case (\cev) while the dashed
  lines are from Case (\lowM). \label{fig:C_rms}}
\end{figure}

The effects of wall transpiration can also be seen in the RMS
velocities (Figure~\ref{fig:C_rms}).
The streamwise RMS velocity component in particular is greatly enhanced by
wall transpiration, increasing by approximately $30\%$ from Case
(\lowM) to Case (\cev).  These observations are consistent with the
results obtained by~\citet{sumitani1995direct} for a channel flow with
an injecting wall and a suction wall.  They showed that turbulent
fluctuations are larger on the injection side as compared with a
channel with an impermeable wall.

\begin{figure}[ht]
\begin{center}
\includegraphics[width=0.6\linewidth]{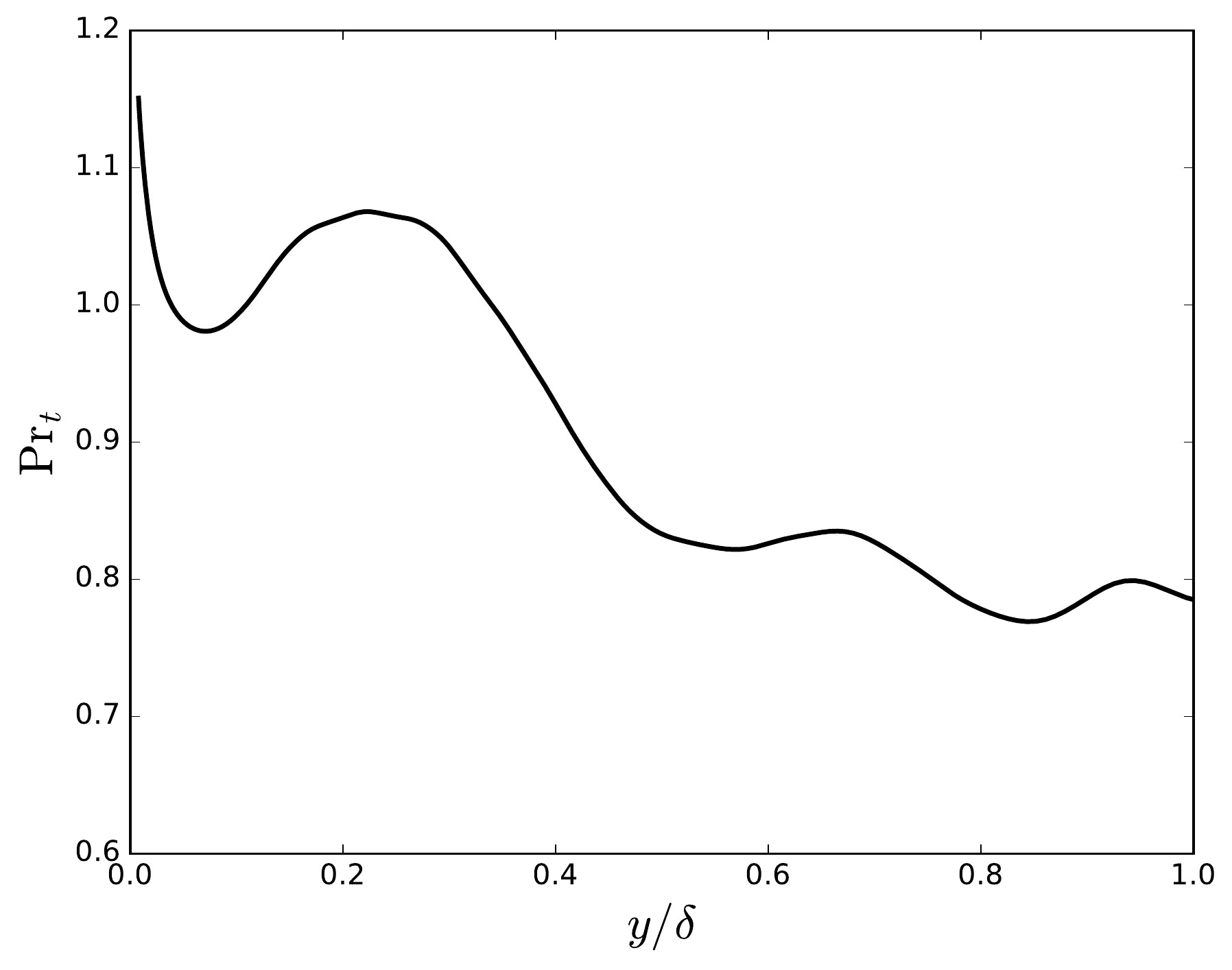}
\end{center}
\caption{Turbulent Prandtl number. \label{fig:C_Prt}}
\end{figure}

To examine the Reynolds heat flux, Figure~\ref{fig:C_Prt} shows the
turbulent Prandtl number:
\begin{equation*}
\Prt
= \frac{\overline{\rho u'' v''} (\partial\fav{T}/\partial y)}
       {\overline{\rho T'' v''}(\partial\fav{u}/\partial y)}.
\end{equation*}
In standard RANS modeling, $\Prt$ is taken to be a constant, usually
$\Prt = 0.9$, although values between $0.6$ and $1.0$ have been used.
Examining the figure, it is clear that, while the standard value of
$\Prt \approx 0.9$ is a reasonable compromise for this case, the true
value varies substantially across the boundary layer, from $\Prt
\approx 0.8$ to greater than $1.1$ near the wall.  Similar values and
trends for $\Prt$ were also observed by~\citet{guarini2000direct}
and~\citet{Pirozzoli2004} in adiabatic, impermeable wall simulations,
indicating that the accuracy of the constant $\Prt$ approximation does
not substantially degrade due to cold wall or blowing effects.

\begin{figure}[htp]
\begin{center}
\subfigure[Near wall region (non-dimensionalized by semi-local scales).]{\includegraphics[width=0.6\linewidth]{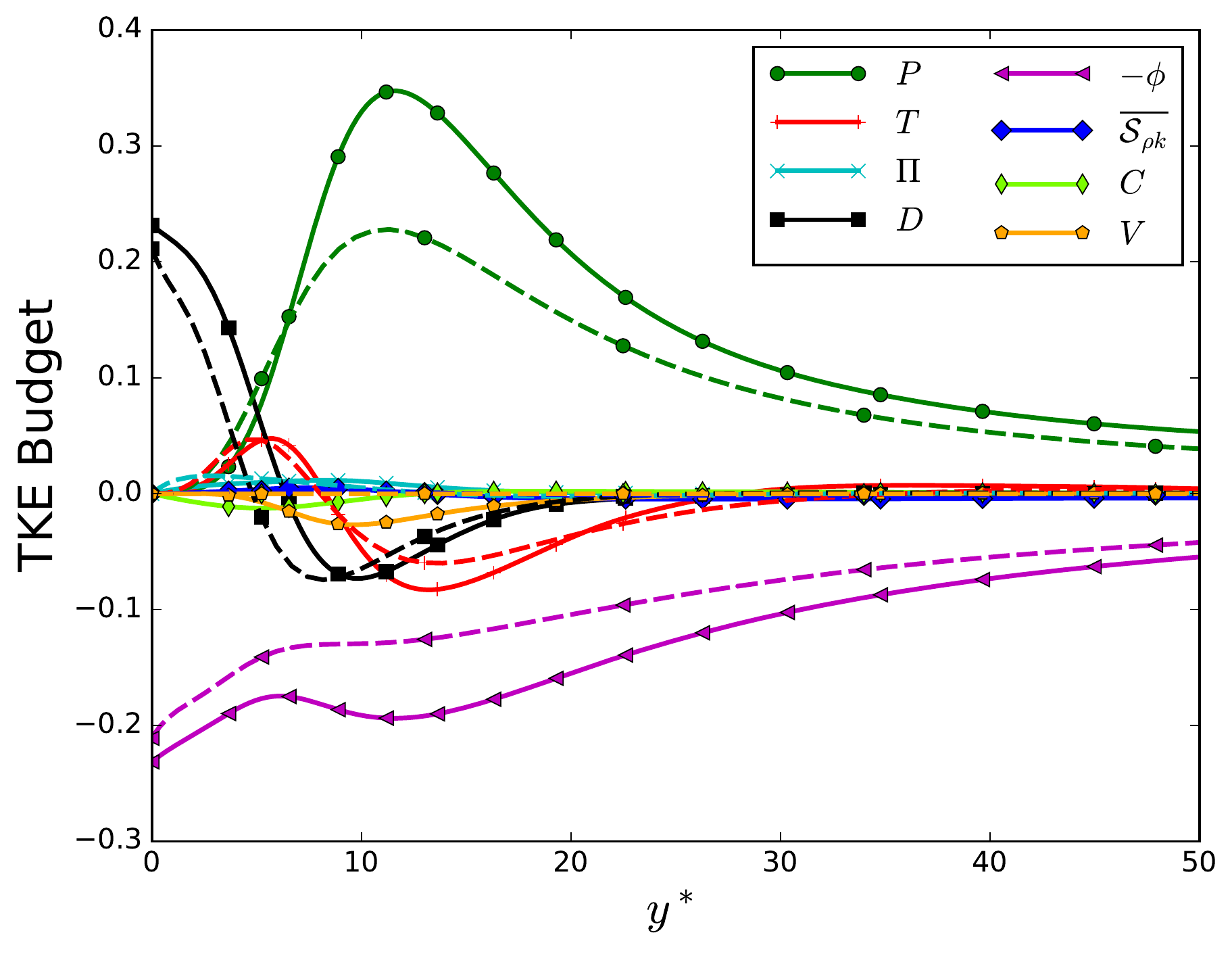}}
\subfigure[Outer region (non-dimensionalized by $\delta$, $\bar{\rho}$, and $u_{\tau^*}$).]{\includegraphics[width=0.6\linewidth]{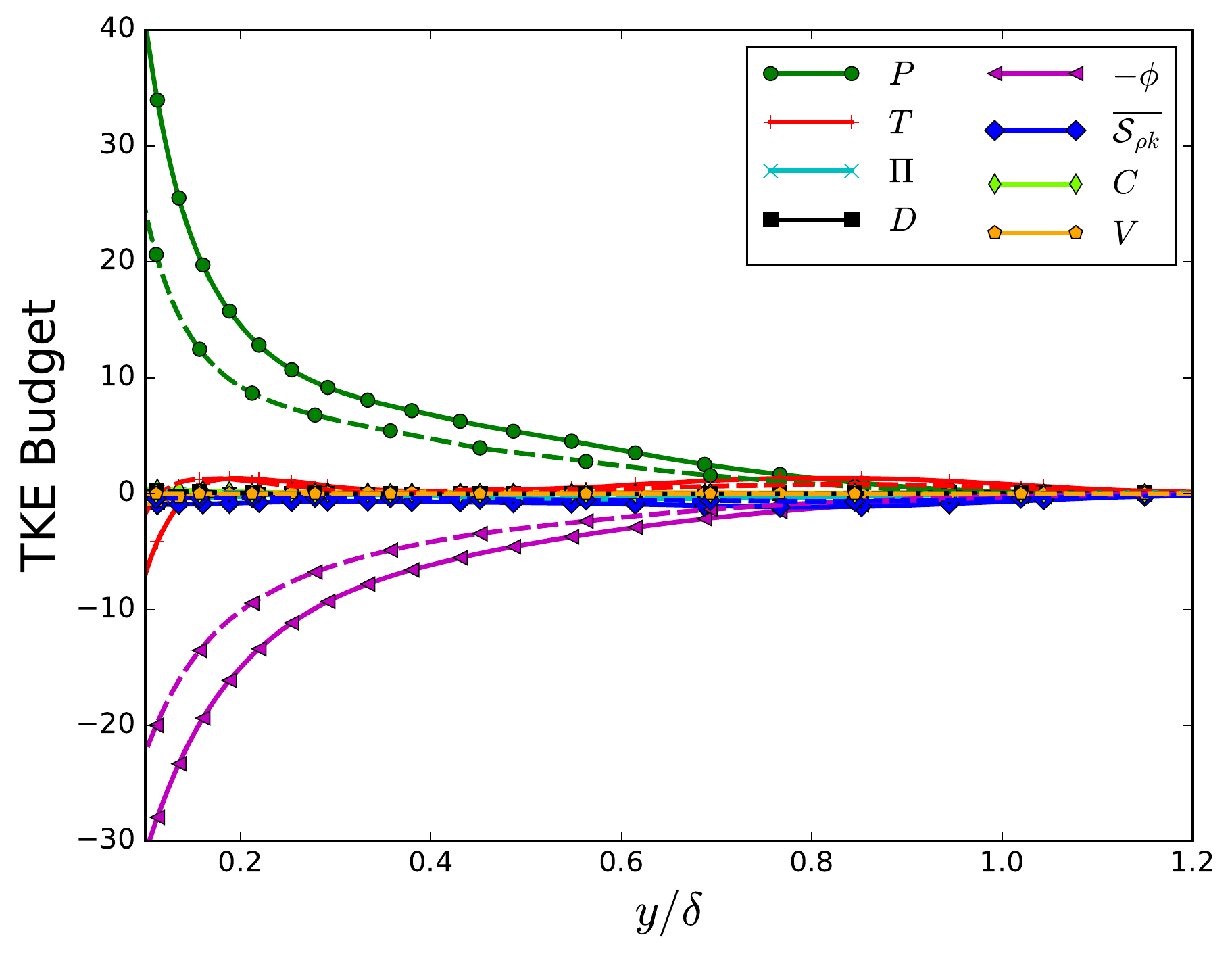}}
\end{center}
\caption{Turbulent kinetic energy budget. Solid lines show results
  from Case (\cev).  Dashed lines show results from Case
  (\lowM). \label{fig:C_tke_budget}}
\end{figure}

Finally, the turbulent kinetic energy budget is shown in Figure~\ref{fig:C_tke_budget}.
Unlike the budget profiles of~\citet{guarini2000direct}, which collapsed
reasonably well with those from the incompressible simulations
of~\citet{spalart1988direct} when non-dimensionalized using $u_{\tau}$
and $\nu_\wall$, the budget for Case (\cev) is substantially different
than that for Case (\lowM).  The near-wall peak in production for Case
(\cev) is almost $50\%$ greater than the peak production in Case
(\lowM), which is consistent with the enhanced Reynolds stress due to
blowing.  The dissipation and turbulent transport are also larger in
magnitude in the near-wall region for Case (\cev) relative to Case
(\lowM).  Further, neither the mean convection $C$ nor the terms
associated with variable density $V$ are entirely negligible.



\section{Conclusions} \label{sec:conclusions}
A new slow growth formulation for DNS of wall-bounded turbulence has
been developed and used to simulate two flows: an essentially
incompressible boundary layer and a transonic boundary layer over a
cooled wall with transpiration.  Like previous slow growth approaches,
the new formulation relies on an assumption that the mean and RMS
quantities evolve slowly relative to the turbulent fluctuations.  This
assumption is used to develop a set of governing equations for the
fast evolution of the turbulent fluctuations subject to forcing from
the slow evolution of the mean and RMS.  After modeling the impact of
the slow evolution in this scenario, one can simulate the fast
evolution at a fixed point in the slow development.

Unlike previous approaches, the present model is developed based on a
temporally evolving boundary layer.  Furthermore, the current approach
is specifically designed to enable calibration and validation of
RANS-based turbulence models for complex boundary layer flows.  It is
formulated to ensure that the slow growth sources that appear in the
RANS equations are closed in terms of the RANS variables.  This avoids
any potential confounding of errors between typical RANS closures and
new modeling required to close the mean slow growth sources.  Further,
the slow growth source terms that arise from the homogenization
procedure are modeled assuming a self-similar evolution of mean and
RMS profiles.  This procedure allows straightforward extensions to
cases involving other physical phenomena such as compressibility,
transpiration and chemical reactions, which have not been addressed in
previous slow growth formulations.

The results show that in the incompressible case the results display
many characteristics associated with typical boundary layer
turbulence.  The mean velocity profile has the typical structure, and
the streamwise RMS velocity peak location and magnitude is consistent
with other simulations.  Other statistics, most notably the total
shear stress and Reynolds shear stress, display notable discrepancies
with spatial simulations.  These discrepancies result from the
difference between the temporal slow growth model and the true slow
evolution of a spatially evolving boundary layer, due both to the
temporal evolution and the slow growth approximations.  This
observation points to the possibility that an improved slow growth
model could reduce this discrepancy and give a better representation
of a spatially developing flow.  While beyond the scope of this paper,
such models have been proposed~\citep{Ulerich2014} and remedy some of
the differences observed here. Nonetheless, despite the mild
discrepancies between the current slow growth formulation and
spatially evolving boundary layers, the slow growth simulations are a
valuable resource for evaluation of RANS models.  Specifically, the
slow growth boundary layer is sufficiently similar to a spatially
evolving one that a model that represents the former should be able to
simulate the later.

Finally, the transonic, cold wall case with wall transpiration shows
that the approach can be straightforwardly extended to problems with
more complex physics.  This capability is significant because it
enables the development of data sets for assessing the validity of
lower fidelity models, namely RANS models, in the presence of these
complicating phenomena. This is particularly useful for calibration
and validation because reliable data for boundary layers with such
complications is often scarce or nonexistent.  Work to further extend
the slow growth capability to treat pressure gradients and reacting
flows is underway.  These capabilities together will enable affordable
DNS of boundary layer flows similar to those observed on vehicles
during atmospheric entry and in other complex systems.

\section*{Acknowledgments}
This material is based in part upon work supported by the Department of
Energy [National Nuclear Security Administration] under Award Number
[DE-FC52-08NA28615].

The authors acknowledge the Texas Advanced Computing Center (TACC) at
The University of Texas at Austin for providing HPC resources that
have contributed to the research results reported within this
paper. URL: \url{http://www.tacc.utexas.edu}

\appendix
\section{An inconsistent slow growth formulation} 
\label{sec:inconsistent}

A straightforward formulation can be obtained by considering a Reynolds
decomposition of the conserved variables. 
Specifically, let
\begin{equation*}
\func{\rho q}{x, y, z, t} = 
  \func{\mean{\rho q}}{y, t_s} +
  \underbrace{ 
   \func{A_{\rho q}}{y,t_s} \func{\fluc{\rho q}_p}{x,y,z,t_f}
  }_{\func{\fluc{\rho q}}{x,y,z,t_f,t_s}},
\end{equation*}
where the mean $\mean{\rho q}$ and amplitude function $A_{\rho q}$ are
assumed to evolve only in slow time.  As in
Section~\ref{sec:construct_model}, to model the slow time derivatives, the
mean and amplitude are assumed to evolve in time in a self-similar
manner:
\begin{align*}
\func{\mean{\rho q}}{t_s, y} &= \func{F_{\rho q}}{y / \Delta(t_s)}, \\
\func{A_{\rho q}}{t_s, y} &= \func{G_{\rho q}}{y / \Delta(t_s)}.
\end{align*}

Then, by an exactly analogous development to that shown in
Section~\ref{sec:construct_model}, the slow growth source for $\rho q$ is
found to be
\begin{equation*}
\Ssd_{\rho q} 
    = y \, \grt(\Delta)
          \left(
             \pp{\mean{\rho q}}{y} 
             + \frac{\fluc{(\rho q)}}{A_{\rho q}} \pp{A_{\rho q}}{y}
          \right).
\end{equation*}
Without specifying $A_{\rho q}$, it is clear that, while the mean of
$\Ssd_{\rho q}$ is closed in terms of the mean flow, the mean of the
slow growth source in the TKE equation cannot be closed without
additional modeling in this formulation.  Thus, the formulation is
discarded in favor of that shown in Section~\ref{sec:construct_model}.
Nonetheless, we have performed simulations using this formulation, and
it leads to similar results to those presented in this work.  This
observation indicates that the results are not highly sensitive to the
choice of whether to apply the Reynolds decomposition to the primitive
or conserved variables.

\section{Analysis of Total Stress} \label{app:total_stress}
In this section, we examine the relationship between the mean
streamwise velocity and the total stress in a spatially evolving
boundary layer, a temporally evolving boundary layer, and the temporal
slow growth model.  In particular, we consider a
zero-pressure-gradient, constant-density boundary layer flow and
analyze the appropriate form of the boundary layer equations for each
case.  The implied behavior of the total stress for the different
cases explains the near wall differences observed in the total stress
profiles shown in Section~\ref{sec:results_lowM}.

\subsection{Spatially Evolving Boundary Layer}
For the spatially evolving case, the mean boundary layer equations can
be written
\begin{gather*}
\pp{u}{x} + \pp{v}{y} = 0, \\
u \pp{u}{x} + v \pp{u}{y} = \pp{\tau}{y},
\end{gather*}
where $x$ is the streamwise direction, $y$ is the wall-normal
direction, $u$ and $v$ are the mean streamwise and wall-normal
velocities, respectively, and $\tau$ is the mean total shear stress.
Assuming that the streamwise velocity normalized by the friction
velocity is only a function of wall-normal distance normalized by the
viscous length scale, i.e., $u(x,y) / u_{\tau}(x) = u^+(y^+)$, one can
derive a relationship between the total shear stress and the velocity.
To begin, note that
\begin{equation*}
u = u_{\tau}(x) \, u^+(y^+)
\quad \Rightarrow \quad
\pp{u}{x} 
= \dd{u_{\tau}}{x} \left( u^+ + y^+ \dd{u^+}{y^+}\right)
= \dd{u_{\tau}}{x} \, \dd{}{y^+} \left( y^+ u^+ \right).
\end{equation*}
Thus, the wall-normal velocity is given by
\begin{equation*}
v(x,y) 
= - \int_0^y \pp{u}{x} d y 
= - \int_0^{y^+} \dd{u_{\tau}}{x} \, \dd{}{y^+} \left( y^+ u^+ \right) \frac{\nu}{u_{\tau}} \, d y^+
= - \, \frac{\nu}{u_{\tau}} \dd{u_{\tau}}{x} y^+ u^+.
\end{equation*}
Using these results to evaluate the convection term in the mean momentum
equation gives
\begin{gather*}
u \pp{u}{x} = u_{\tau} \dd{u_{\tau}}{x} \, u^+ \dd{}{y^+} \left( y^+ u^+ \right), \\
v \pp{u}{y} 
= - \dd{u_{\tau}}{x} y^+ u^+ \frac{\nu}{u_{\tau}} u_{\tau} \dd{u^+}{y^+} \frac{u_{\tau}}{\nu}
= - u_{\tau} \dd{u_{\tau}}{x} y^+ u^+ \dd{u^+}{y^+}.
\end{gather*}
Thus,
\begin{equation*}
u \pp{u}{x} + v \pp{u}{y} 
= 
u_{\tau} \dd{u_{\tau}}{x} \left( u^+ \dd{}{y^+} \left( y^+ u^+ \right) - y^+ u^+ \dd{u^+}{y^+} \right)
=
u_{\tau} \dd{u_{\tau}}{x} (u^{+})^2.
\end{equation*}
Substituting into the mean momentum equation gives
\begin{equation*}
u_{\tau} \dd{u_{\tau}}{x} (u^+)^2
=
\pp{\tau}{y}.
\end{equation*}
Thus,
\begin{equation*}
\tau - \tau_{\wall} 
= 
\int_0^y \pp{\tau}{y} dy
= 
u_{\tau} \dd{u_{\tau}}{x} \int_0^y (u^+)^2 \diff y.
\end{equation*}
Finally, non-dimensionalizing by  $\nu$ and $u_{\tau}$ gives
\begin{equation*}
\frac{\tau}{\tau_{\wall}}
=
1 + \left( \frac{\nu}{u_{\tau}^2}  \dd{u_{\tau}}{x} \right) \int_{0}^{y^+} (u^+)^2 \, \diff y^+.
\end{equation*}

\subsection{Temporally Evolving Boundary Layer}
In the temporally evolving case, the flow is necessarily homogeneous
in the streamwise direction.  Conservation of mass plus the no slip
condition implies that $v = 0$.  Thus, the boundary layer
equations reduce to
\begin{equation*}
\pp{u}{t} = \pp{\tau}{y}.
\end{equation*}
As in the spatially evolving case, we assume that $u^+$ is a universal
function of $y^+$ only.  Then,
\begin{equation*}
\pp{u}{t} 
=
\dd{u_{\tau}}{t} u^+ + y^+ \dd{u^+}{y^+} \dd{u_{\tau}}{t}
=
\dd{u_{\tau}}{t} \dd{\,(u^+ y^+)}{y^+}.
\end{equation*}
Substituting this result into mean momentum and integrating gives
\begin{equation*}
\tau - \tau_\wall 
= 
\int_0^y \pp{\tau}{y} \diff y
= 
\frac{\nu}{u_{\tau}} \dd{u_{\tau}}{t} \, (u^+ y^+).
\end{equation*}
Thus, non-dimensionalizing using $\nu$ and $u_{\tau}$ gives
\begin{equation*}
\frac{\tau}{\tau_\wall}
= 
1 + \left( \frac{\nu}{u_{\tau}^3} \dd{u_{\tau}}{t} \right) (u^+ y^+).
\end{equation*}

\subsection{Temporal Slow Growth Boundary Layer}
The temporal slow growth solution is also homogeneous in the
streamwise direction, which leads to $v = 0$, as in the
temporally evolving case.  In addition, the flow is statistically
stationary by design.  Thus, the boundary layer equations become
\begin{equation*}
0 = \pp{\tau}{y} + \overline{S_{u}},
\end{equation*}
where
\begin{equation*}
\overline{S_u} = y \, \grt(\Delta) \pp{u}{y}.
\end{equation*}
Thus,
\begin{equation*}
\tau = \tau_\wall - \grt(\Delta) \left[ u y - \int_{0}^{y} u \diff y \right],
\end{equation*}
and
\begin{equation*}
\frac{\tau}{\tau_\wall}
=
1 - \left( \frac{\nu}{u_{\tau}^2} \grt(\Delta) \right) \left[ u^+ y^+ - \int_{0}^{y^+} u^+ \diff y^+ \right].
\end{equation*}

\section{An Extended Van Driest Transformation} \label{app:evd}
We construct an extension of the van Driest transformation that
accounts for the effects of wall transpiration and wall cooling.  The
van Driest transformation~\citep{vanDriestTrans} is derived using the following
relationship between the compressible mean velocity ($\tilde{u}$) and
the incompressible mean velocity ($\bar{u}_{\inc}$):
\begin{equation}
\dd{\tilde{u}^+}{y^+} = \frac{(\bar{\rho} / \bar{\rho}_\wall)^{1/2}}{\kappa y^+} = \left(\frac{\bar{\rho}}{\bar{\rho}_\wall} \right)^{1/2} \dd{\bar{u}^+_{\inc}}{y^+},
\label{eqn:vd0}
\end{equation}
which is valid in the log layer.  As pointed out
by~\citet{huang1994van}, in the viscous sublayer,~\eqref{eqn:vd0} is
incorrect.  Instead, in the sublayer, the correct relationship is
\begin{equation*}
\dd{\tilde{u}^+}{y^+} = \frac{\mu_\wall}{\mu} \dd{\bar{u}^+_{\inc}}{y^+}.
\end{equation*}
The van Driest transformation is derived by
integrating~\eqref{eqn:vd0} starting at the wall, without any
correction for the viscous sublayer.  Strictly speaking, this
procedure is always incorrect, but as long as the temperature does not
vary dramatically in the viscous sublayer, the difference between
$(\mu_\wall / \mu)$ and $(\bar{\rho} /\bar{\rho}_\wall)^{1/2}$ is not
large, and the resulting transformed velocity profile agrees well with
incompressible results~\citep{White_1991_Viscous_Flow}.  However, when
the temperature variation in the sublayer is large, as it is for a
cold wall as shown in Section~\ref{sec:results_highM}, the error due
to the sublayer is large enough that the collapse between the
transformed profile and the incompressible results is quite poor.

To remedy this error,~\citet{huang1994van} proposed a blending between
the viscous sublayer and log layer results based on an assumed mixing
length.  Here, this approach is extended to include the effect of wall
transpiration.  The primary effect of wall
transpiration is that the wall normal mean convection term in the mean
momentum equation is no longer negligible near the wall.  Thus, rather
than containing only the viscous and Reynolds shear stresses, the
total shear stress contains a contribution from wall-normal
convection.  Using the boundary layer form of the slow growth mean momentum
equation, one can show that
\begin{equation*}
\bar{\rho} \tilde{u} \tilde{v}
=
\tau - \tau_{\wall} + \int_{0}^{y} \overline{\Ssd_{\rho u}} \diff y.
\end{equation*}
An analogous development can be done for the spatially developing
case.  For brevity, this analysis is not shown since only the slow
growth version is used here.

Since $\tau = \mu \partial \tilde{u} / \partial y -
\overline{\rho u'' v''}$ (where $\mu$ is the mean viscosity and we
have neglected the viscosity/velocity gradient correlation), the wall
shear stress can be written as
\begin{equation}
\tau_\wall
= 
- \bar{\rho} \tilde{u} \tilde{v} + \int_{0}^{y} \overline{\Ssd_{\rho u}} \diff y 
+ \mu \pp{\tilde{u}}{y} - \overline{\rho u'' v''}.
\label{eqn:evd_tauw}
\end{equation}
To simplify notation, the convection and slow growth source terms can
be grouped together.  Note that
\begin{equation*}
\int_{0}^{y} \overline{\Ssd_{\rho u}}
=
\int_{0}^{y} y \grt(\Delta) \pp{\overline{\rho u}}{y}
=
y \grt(\Delta) \overline{\rho u} - \grt(\Delta) \int_{0}^{y} \overline{\rho u}.
\end{equation*}
Thus,
\begin{equation*}
\bar{\rho} \tilde{u} \tilde{v} - \int_{0}^{y} \overline{\Ssd_{\rho u}} \diff y
=
\bar{\rho} \tilde{u} \left( \tilde{v} - y \grt(\Delta) + \grt(\Delta) \int_{0}^{y} \frac{\overline{\rho u}(\eta)}{\overline{\rho u}(y)} \, \diff \eta \right).
\end{equation*}
Then, let
\begin{equation*}
\vmod =  \tilde{v} - y \grt(\Delta) + \grt(\Delta) \int_{0}^{y} \frac{\overline{\rho u}(\eta)}{\overline{\rho u}(y)} \, \diff \eta.
\end{equation*}
With this notation,~\eqref{eqn:evd_tauw} can be rewritten as
\begin{equation}
\tau_\wall
= 
- \bar{\rho} \tilde{u} \vmod
+ \mu \pp{\tilde{u}}{y} - \overline{\rho u'' v''}.
\label{eqn:evd_tauw_simple}
\end{equation}

To continue, we use a mixing length model for the Reynolds stress:
\begin{equation*}
- \overline{\rho v'' u''} = \bar{\rho} \ell^2 \left( \pp{\tilde{u}}{y} \right)^2.
\end{equation*}
Then,~\eqref{eqn:evd_tauw_simple} can be rewritten as
\begin{equation*}
\tau_\wall = - \bar{\rho} \vmod \tilde{u} + \mu \pp{\tilde{u}}{y} + \bar{\rho} \ell^2 \left( \pp{\tilde{u}}{y} \right)^2.
\end{equation*}
Solving this quadratic for $\partial \tilde{u} / \partial y$, one obtains
\begin{equation*}
\pp{\tilde{u}}{y}
=
\frac{2 \left( \bar{\rho} \vmod \tilde{u} + \tau_\wall \right)}
     {\mu + \sqrt{ \strut \mu^2 + 4 \bar{\rho} \ell^2 \left( \bar{\rho} \vmod \tilde{u} + \tau_\wall \right)}}\,.
\end{equation*}
Non-dimensionalizing this result using $\rho_\wall$, $\mu_\wall$, and
$u_{\tau}$ gives
\begin{equation}
\pp{\tilde{u}^+}{y^+} 
=
\frac{2 \left( \hat{\rho} \vmod^+ \tilde{u}^+ + 1 \right)}
     {\hat{\mu}^2 + \sqrt{\strut \hat{\mu}^2 + 4 \hat{\rho} (\ell^+)^2 \left( \hat{\rho} \vmod^+ \tilde{u}^+ + 1 \right)}}\,,
\label{eqn:dudy_comp}
\end{equation}
where $\hat{\rho} = \bar{\rho}/\rho_\wall$, $\hat{\mu} =
\mu/\mu_\wall$, and $\ell^+ = \rho_\wall u_{\tau} \ell /
\mu_\wall$.  In the incompressible, non-blowing wall case, this result
simplifies to
\begin{equation}
\pp{\bar{u}_{\inc}^+}{y^+}
=
\frac{2}{1 + \sqrt{ 1  + 4 (\ell_{\inc}^+)^2}},
\label{eqn:dudy_incomp}
\end{equation}
where $\ell_{\inc}$ is the mixing length for the incompressible case.

The extended van Driest transformation is obtained by requiring that
the nondimensional transformed velocity $\ueff^+$ has the same profile as
the incompressible velocity.  That is,
\begin{equation*}
\ueff^+( \tilde{u}^+ (y^+) ) = \bar{u}^+_{\inc}(y^+),
\end{equation*}
which implies that
\begin{equation*}
\dd{\ueff^+}{\tilde{u}^+} \, \dd{\tilde{u}^+}{y^+} = \dd{\bar{u}_{\inc}^+}{y^+}.
\end{equation*}
Thus,
\begin{equation}
\ueff^+(w^+)
= \int_{0}^{w^+} \dd{\ueff^+}{\tilde{u}^+} \, \diff \tilde{u}^+
= \int_{0}^{w^+} \frac{ \diff \bar{u}_{\inc}^+ / \diff y^+}{ \diff \tilde{u}^+ / \diff y^+} \, \diff \tilde{u}^+.
\label{eqn:transform0}
\end{equation}
Substituting~\eqref{eqn:dudy_comp} and~\eqref{eqn:dudy_incomp}
into~\eqref{eqn:transform0} gives
\begin{equation}
\ueff^+(w^+)
=
\bigintss_{0}^{w^+}
\frac{\hat{\mu} + \sqrt{\strut \hat{\mu}^2 + 4 \hat{\rho} (\ell^+)^2 \left( \hat{\rho} \vmod^+ \tilde{u}^+ + 1 \right)}}
     {(\hat{\rho} \vmod^+ \tilde{u}^+ + 1) \left( \sqrt{1 + 4 (\ell_{\inc}^+)^2} + 1 \right)} \, \diff \tilde{u}^+.
\label{eqn:transform1}
\end{equation}
To complete the transformation, one must define the mixing length.  We
use the van Driest damping~\citep{vanDriestDamp} function:
\begin{gather*}
\ell^+ = \kappa y^+ \left( 1 - \exp(-y^* / A^+) \right), \\
\ell_{\inc}^+ = \kappa y^+ \left( 1 - \exp(-y^+ / A^+) \right),
\end{gather*}
where the wall-distance normalized by the semi-local viscous length is
used in the compressible case.  We take typical
values of the parameters: $\kappa = 0.41$ and $A^+ = 25.51$.

At this point, if profiles for $\hat{\rho}$, $\hat{\mu}$, $\vmod^+$
and $\tilde{u}^+$ are available,~\eqref{eqn:transform1} allows
computation of the equivalent incompressible profile.  Thus, this form
is appropriate for data analysis, and it is used for this purpose in
Section~\ref{sec:results_highM}.  However, it does not give a closed form
for modeling a compressible profile.  For this task, a model
temperature profile is required.  See~\citet{huang1994van} for an
example.

To conclude, we examine how the extended transformation compares to
existing transformations.  When $\vmod^+$ is negligible (i.e., no wall
transpiration or slow growth), the transformation reduces to the
method of~\citet{huang1994van}, which itself reduces to the standard van
Driest transformation outside the viscous sublayer.  Alternatively, for
the incompressible case with wall transpiration, the extended
transformation in the log layer reduces to the log layer correction
for injection effects derived by \citet{stevenson1963law}.

\bibliographystyle{jfm}
\bibliography{paper}

\end{document}